# Early-growing Supermassive Black Holes Strengthen Bars and Boxy/Peanut Bulges

Vance Wheeler[1,2], Monica Valluri[1], Leandro Beraldo e Silva[1], Shashank Dattathri[1,3], and Victor P. Debattista[4]
[1] Department of Astronomy, University of Michigan, 1085 S. University Avenue, Ann Arbor, MI 48109, USA; vanwheel@umich.edu
[2] Department of Physics, University of Chicago, Chicago, IL 60637, USA
[3] Department of Astronomy, Yale University, P.O. Box 208101, New Haven, CT 06520, USA
[4] Jeremiah Horrocks Institute, University of Central Lancashire, Preston PR1 2HE, UK

## Abstract

Using $N$-body simulations, we explore the effects of growing a supermassive black hole (SMBH) prior to or during the formation of a stellar bar. Keeping the final mass and growth rate of the SMBH fixed, we show that if it is introduced before or while the bar is still growing, the SMBH does not cause a decrease in bar amplitude. Rather, in most cases, it is strengthened. In addition, an early-growing SMBH always either decreases the buckling amplitude, delays buckling, or both. This weakening of buckling is caused by an increase in the disk vertical velocity dispersion at radii well beyond the nominal black hole sphere of influence. While we find considerable stochasticity and sensitivity to initial conditions, the only case where the SMBH causes a decrease in bar amplitude is when it is introduced after the bar has attained a steady state. In this case, we confirm previous findings that the decrease in bar strength is a result of scattering of bar-supporting orbits with small pericenter radii. By heating the inner disk both radially and vertically, an early-growing SMBH increases the fraction of stars that can be captured by the inner Lindblad resonance (ILR) and the vertical ILR, thereby strengthening both the bar and the boxy-peanut-shaped bulge. Using orbital frequency analysis of star particles, we show that when an SMBH is introduced early and the bar forms around it, the bar is populated by different families of regular bar-supporting orbits than when the bar forms without an SMBH.

*Unified Astronomy Thesaurus concepts:* Galaxy evolution (594); Barred spiral galaxies (136); Supermassive black holes (1663); Galaxy bulges (578); Galaxy bars (2364); Galaxy nuclei (609); N-body simulations (1083); Orbital resonances (1181)

## 1. Introduction

One of the most consequential astronomical discoveries of the past 25 yr is that most massive galaxies contain supermassive black holes (SMBHs) in their nuclei, and that the large-scale properties of host galaxies and the masses of nuclear SMBHs are linked through several scaling relations such as those between black hole mass $M_{\rm BH}$ and central stellar velocity dispersion $\sigma$ (the $M_{\rm BH}$–$\sigma$ relation; e.g., Ferrarese & Merritt 2000; Gebhardt et al. 2000; Merritt & Ferrarese 2001; Gebhardt et al. 2003; Gültekin et al. 2009; McConnell & Ma 2013; Saglia et al. 2016), $M_{\rm BH}$ and bulge (spheroid) stellar mass or luminosity (e.g., Häring & Rix 2004; Gültekin et al. 2009; Scott et al. 2013), $M_{\rm BH}$ and stellar light concentration (Graham et al. 2001; Savorgnan et al. 2013), as well as correlations with several other galaxy properties (for recent compilations and reviews see, e.g., Kormendy & Ho 2013; Saglia et al. 2016). These scaling relations are widely considered to be evidence that, despite constituting only ~0.2% of the mass of their hosts, SMBHs (via active galactic nuclei, AGNs) influence the growth of the galaxies on scales that are orders of magnitude larger than the black hole sphere of influence, $r_{\rm BH}$.[5]

While the origin of these scaling relations is still actively debated, the view that the scaling relations are evidence for tight coevolution between SMBHs and their host galaxies mediated by AGN feedback (Fabian 2012; King 2014) is being replaced by the view that SMBH growth may also be driven by the averaging of BH and galaxy properties via hierarchical merging (Jahnke & Macciò 2011) and by secular evolution in disk galaxies.

Nonaxisymmetric disk structures like bars and spirals have long been considered important drivers of secular evolution in disk galaxies, which for instance can lead to the formation of pseudobulges (Kormendy & Kennicutt 2004) as well as the boxy-peanut-shaped bulges that are observed in at least half of all edge-on disk galaxies (Lütticke et al. 2000) and probably exist in most massive barred galaxies (Erwin & Debattista 2017). Simulations show that bars (especially nested bars) can facilitate the outward transport of angular momentum and the inflow of gas and stars (Shlosman et al. 1989). In the presence of gas, which experiences shocks and undergoes clumping, a diverse range of nonaxisymmetric morphologies (spirals, rings, and bars; Hopkins & Quataert 2010, 2011) enables the outward transport of angular momentum and the inward flow of mass over a wide range of physical scales (from kiloparsec to subparsec scales) down to the accretion disks that ultimately fuel the central black holes.

Several authors have found that black holes in late-type galaxies, especially those with bars and pseudobulges, fall below the black hole scaling relations for higher-mass elliptical galaxies and disk galaxies with classical bulges (e.g., Graham 2008; Greene et al. 2010; Graham et al. 2011; Kormendy et al. 2011; Graham & Scott 2015). However, in a recent study of 66 local AGNs, Bennert et al. (2021) used high-spatial-resolution Hubble Space Telescope imaging, ground-based high-resolution near-infrared imaging with adaptive optics and sophisticated modeling of surface photometry to obtain accurate morphological classifications and characterization of the spheroids of the AGN host galaxies. They combined these data with high spatial resolution spectroscopy obtained with Keck/LRIS (Harris et al. 2012) to determine the central stellar velocity dispersions in all of the galaxies in a consistent manner and used single-epoch reverberation mapping to determine black hole masses for the entire sample. Their analysis found no dependence on the morphological type of the host, presence of either a pseudo or classical bulge, or the presence of a bar for any of the three black hole scaling relations that they examined, supporting the idea that secular evolution in disks remains a viable black hole growth mechanism. However, since this latter study focused only on active galaxies while the former studies focused on quiescent galaxies, it is unclear whether there is a discrepancy.

---
[5] The radius within which the mass of the SMBH is equal or greater to that of the stellar component. The definition based on velocity dispersion within the effective radius was not used because effective radius is difficult to determine and changes significantly as the bulge grows. Effective radius is also undefined at times before the bulge has formed.

The enthusiasm for bar-mediated SMBH growth was dampened in part by the fairly weak observational evidence for correlations between the existence of AGNs and bars (Cisternas et al. 2013) and in part as result of a series of N-body simulations showing that the growth of central black holes weakened or even destroyed stellar bars (e.g., Hasan et al. 1993; Norman et al. 1996; Shen & Sellwood 2004; Athanassoula et al. 2005; Hozumi & Hernquist 2005; Du et al. 2017). These N-body simulations showed that when a central point mass representing an SMBH of realistic bulge mass fraction (∼0.10%–0.20%) was introduced into a bar that had reached a steady state (i.e., the bar strength was no longer changing), it weakened the bar. While unrealistically large SMBH masses of 2%–5% were required to destroy large-scale bars, realistic SMBH mass fractions of ∼0.20% could weaken bars and even destroy nuclear bars (Du et al. 2017). The weakening and destruction of the bar has been primarily attributed to chaotic scattering of a significant fraction of the bar-supporting centrophilic orbits by the compact central black hole (Hasan et al. 1993; Merritt & Valluri 1996; Norman et al. 1996; Shen & Sellwood 2004). While the details of the simulations varied, all of these authors introduced the point mass representing the SMBH after the bar had attained steady state and found similar results regarding bar weakening and/or destruction. Shen & Sellwood (2004) noted that their bars buckled prior to the introduction of the central mass concentration (CMC), while Hozumi & Hernquist (2005) studied a 2D model, where buckling is not possible, and the others do not comment on the prior evolution of their bars.

Seed black holes are expected to form at high redshifts in compact galaxies that evolve to form the galactic nuclei of most present-day galaxies (Volonteri 2010). Most present-day disk galaxies probably form around such galactic nuclei, and there is evidence that since at least $z = 2$, a significant fraction of SMBH growth has occurred primarily in disk galaxies probably driven by secular evolution rather than mergers (Gabor et al. 2009; Georgakakis et al. 2009; Cisternas et al. 2011; Schawinski et al. 2011; Kocevski et al. 2012; Donley et al. 2018).

In the local Universe, ∼65% of disks have stellar bars (e.g., Knapen 1999; Eskridge et al. 2002; Marinova & Jogee 2007). This fraction drops to ∼20% by $z = 0.84$ (Sheth et al. 2008). A study of AGNs and quiescent barred galaxies shows that when matched for stellar mass, disks with an AGN have a slightly higher bar fraction than inactive disks (Cisternas et al. 2015); although, the differences between the fraction of bars in active and inactive disks decreases with increasing redshift and could be consistent with no difference. However, since bars are generally long-lived (surviving many gigayears in simulations) but AGN duty cycles are short ($<10^8$ yr), the absence of a strong correlation between the presence of a bar and the presence of an AGN is unlikely to be a strong indicator of the importance (or not) of bars.

Boxy-peanut/X (hereafter BP/X)–shaped bulges are so called because of the easily identifiable eponymous shape they present when the disk is viewed edge-on and the bar major axis lies between ∼30° and 90° to the line of sight. Simulations have shown that disks easily form bars and BP/X bulges from a variety of initial conditions. Early simulations showed that BP/X bulges can form following a buckling event in a bar (e.g., Combes et al. 1990; Pfenniger & Friedli 1991; Raha et al. 1991). Bending and buckling instabilities were first described by Toomre (1966) for an idealized infinite, uniform density, thin sheet and further investigated by simulations and analytic work in increasingly more complex and realistic stellar distributions (e.g., Araki 1985; Raha et al. 1991; Merritt & Sellwood 1994; Sellwood & Merritt 1994; Debattista et al. 2017; Collier 2020). Bar buckling, which occurs over a short interval of time, results from an asymmetric bending of the bar out of the disk midplane, with the inner portion moving upward (downward) and the outer portions moving downward (upward). This instability is believed to arise because the radial stellar velocity dispersion ($\sigma_R$; dispersion along the length of the bar) increases as the bar lengthens and strengthens, but the vertical velocity dispersion ($\sigma_z$) is not significantly altered by bar growth. As a result, a small vertical displacement of the bar mid-section relative to the disk plane causes the highly radially anisotropic bar-supporting orbits to speed along a curve experiencing a (centrifugal) force perpendicular to the disk along the radius of the curve, further increasing the displacement in the same direction (Raha et al. 1991; Merritt & Sellwood 1994). Buckling results in a redistribution of kinetic energy from the radial direction to the vertical direction. The resultant vertical heating dramatically thickens the bar, while reducing its radial extent. The asymmetric buckling event itself is short lived and produces a thicker bar that is approximately symmetric about the midplane (and may have a BP/X shape). Simulations show that although buckling heats the disk and weakens a bar (Martinez-Valpuesta & Shlosman 2004; Martinez-Valpuesta et al. 2006), bars may continue to grow after a buckling event and may even undergo subsequent buckling events (e.g., Martinez-Valpuesta et al. 2006; Collier 2020). Observational evidence for ongoing buckling has also been reported (Erwin & Debattista 2016) and confirms that buckling does occur in real galaxies and is short lived like in simulations.

While it has long been thought that the buckling event itself is responsible for the formation of the BP/X structure associated with bars, there is growing evidence that orbital resonances could play an important role in producing and enhancing these structures (e.g., Quillen 2002; Quillen et al. 2014; Sellwood & Gerhard 2020). In this resonant sweeping scenario, orbits are still altered by interaction with the resonance, which causes them to be elevated to high $|z|$, but do not become trapped in the resonance permanently, implying that nonresonant orbits primarily contribute to the overall bulge structure (Quillen et al. 2014; Sellwood & Gerhard 2020).

To our knowledge, no previously published works describe how preexisting SMBHs or the early growth of a black hole

(e.g., during bar formation) affect the structure of a bar and its associated BP/X bulge. The aim of this work is to examine how the early growth of SMBHs, either before or coeval with bar formation, can affect the structure of the bar, including the boxy/peanut-shaped bulge, present in most massive barred galaxies. In this paper we explore the interaction and coevolution of the SMBH with the bar using a suite of pure N-body model disk galaxies that naturally form bars susceptible to buckling; we vary only the time at which we begin to grow an SMBH relative to the formation time of the bar in the control model (which does not include an SMBH). Our SMBHs begin to grow at all stages of bar evolution: from the very start in the initially axisymmetric model, at various times throughout bar formation, growth, and buckling, and finally after buckling. In all cases, the growth rate and final mass of the SMBH are kept fixed.

In Section 2 we describe the initial conditions, the SMBH growth parameters, and the N-body simulation method and parameters. In Section 3 we describe the effects of the SMBH on the bar and its buckling and also discuss the sensitivity of our results to small changes in initial conditions. In Section 4 we give a brief description of bar buckling, its dependence on stellar velocity anisotropy and show how and why the SMBH alters the buckling behavior. In Section 5 we explore the importance of resonances in the formation of the BP/X-shaped bulge, and the effect of the SMBH on the trapping of stars into resonances. Finally we discuss a few implications of this work in Section 6 and summarize our results in Section 7. Appendix A describes various numerical tests and additional simulations (including with other initial conditions) that we carried out to validate our results. In Appendix B we provide a full list of input parameters used to run our models.

## 2. N-body Models

### 2.1. Simulation Method and Initial Conditions

We use the grid-based N-body simulation package GALAXY[6] (Sellwood 2014) to simulate the growth of point masses representing an SMBH at various stages in the formation of the bar. All models discussed in this paper were evolved from initial conditions used in previous works (Debattista et al. 2017, 2020; Anderson et al. 2022) and were generated using GALACTICS (Kuijken & Dubinski 1995; Widrow & Dubinski 2005; Widrow et al. 2008). These models consist of exponential disks within a modified Navarro–Frenk–White (NFW; Navarro et al. 1996) live dark matter halo. For all initial conditions the spherical halo density distribution $\rho(r)$ is described by

$$\rho(r) = \frac{2^{2-\gamma}\sigma_h^2}{4\pi a_h^2} \frac{C(r)}{(r/a_h)^\gamma (1 + r/a_h)^{3-\gamma}}, \quad (1)$$

where $\sigma_h$ characterizes the halo velocity dispersion, and $a_h$ characterizes the halo scale radius. $C(r)$ is a function that smoothly truncates the model at a finite radius (Widrow et al. 2008):

$$C(r) = \frac{1}{2}\,\mathrm{erfc}\left(\frac{r - r_h}{\sqrt{2}\,\delta r_h}\right). \quad (2)$$

When $\gamma = 1$ and $r_h \to \infty$, Equation (1) is exactly the NFW distribution. For all models considered in this paper, the halo parameters are $\sigma_h = 400$ km s$^{-1}$, $a_h = 16.7$ kpc, $\gamma = 0.873$, $r_h = 100$ kpc, and $\delta r_h = 25$ kpc (Debattista et al. 2020). The live dark matter halo consists of $4 \times 10^6$ particles of mass $\simeq 1.7 \times 10^5 M_\odot$ each.

The disk has an exponential radial density profile and an isothermal vertical density profile described by

$$\Sigma(R, z) = \Sigma_0 \exp(-R/R_d)\ \mathrm{sech}^2(z/z_d) \quad (3)$$

where $R_d$ ($z_d$) is the disk scale length (height), set to 2.4 kpc (0.3 kpc; Debattista et al. 2020). The initial kinematics of the disk are such that the radial velocity dispersion decreases exponentially as

$$\sigma^2(R) = \sigma_{R0}^2 \exp(-R/R_\sigma), \quad (4)$$

in which $R_\sigma$ is fixed at 2.5 kpc and $\sigma_{R0}$ is the disk's central radial velocity dispersion (Debattista et al. 2020). The disk consists of $6 \times 10^6$ equal mass particles contributing to a total disk[7] mass of $\simeq 5.37 \times 10^{10} M_\odot$.

We assign the disk particles to a cylindrical polar grid nested within a larger spherical grid to which the halo particles are assigned. These grids share an origin that is relocated to the disk particle centroid at regular intervals to ensure the greatest spatial resolution at the central region of greatest density. This model does not contain a classical bulge component, and only forms a boxy or peanut-shaped bulge as a result of secular evolution.

The results in the main body of this work are evolved from the initial conditions for Model 2 from Debattista et al. (2020), for which $\sigma_{R0} = 128$ km s$^{-1}$, and which is our principal control model (Model C). In order to evaluate the robustness of our results against stochastic effects, in Section 3.3 we present results from three azimuthally scrambled versions of Model C, and in Appendix A we also briefly consider results from two additional models from Debattista et al. (2020)—their Model 1 and Model 3 (also known, respectively, as D5 and D2 in Debattista et al. 2017). These models differ by having a value of $\sigma_{R0} = 90$ km s$^{-1}$ and $\sigma_{R0} = 165$ km s$^{-1}$, respectively, with all other parameters identical to the control model.

### 2.2. Growing the SMBH

The SMBH is represented as a Plummer potential using a built-in function in GALAXY and is initialized with a small nonzero SMBH mass. The mass of the SMBH ($M_{\mathrm{BH}}$) grows as a function of time to a final mass $M_{\mathrm{fin}}$ according to the equation:

$$M_{\mathrm{BH}}(t) = \begin{cases} M_{\mathrm{fin}}(0.02 + 0.98\sin^2(\pi\tau/2) & 0 \leqslant \tau \leqslant 1 \\ M_{\mathrm{fin}} & \tau > 1 \end{cases} \quad (5)$$

where $\tau = t/t_{\mathrm{grow}}$, and $t_{\mathrm{grow}}$ is the timescale over which the SMBH grows.

Since this is a pure N-body simulation, accretion by the SMBH is not modeled. Rather, our SMBH grows strictly due to an artificial deepening of the Plummer potential as given by Equation (5). The final mass of the SMBH in all cases is $0.0014\,M_{\mathrm{disk}}$, which is comparable to the black hole mass

---

[6] Publicly available at http://www.physics.rutgers.edu/galaxy/.

[7] Note that the disk total mass and individual disk particle mass were misquoted in Debattista et al. (2020).

**Table 1**
Simulation Parameters for Models

| | Parameter | Value |
|---|---|---|
| General | Base Time Step | $3.784 \times 10^{-2}$ Myr |
| | Softening Length | 50 pc |
| | Disk Mass ($M_{disk}$) | $5.37 \times 10^{10}\ M_\odot$ |
| | Halo Mass | $6.77 \times 10^{11}\ M_\odot$ |
| SMBH | Final Mass ($M_{fin}$) | $0.0014\ M_{disk}$ |
| | Initial Mass | $0.02\ M_{fin}$ |
| | Growth Period ($t_{grow}$) | 378 Myr |
| | Softening Length | 33.33 pc |

**Note.** "General" parameters apply to all models, while "SMBH" parameters are common to all models containing an SMBH. A comprehensive list of simulation parameters may be found in Appendix B.

**Table 2**
List of Principal Set of Models Considered in This Work

| SMBH Growth Epoch | Model Name | SMBH Inserted |
|---|---|---|
| "Control"—No SMBH | | |
| | C | ... |
| Before bar formation | | |
| | $BF_0$ | 0 Gyr |
| Before bar buckling | | |
| | $BB_1$ | 0.575 Gyr |
| | $BB_2$ | 1.150 Gyr |
| | $BB_3$ | 1.877 Gyr |
| After bar buckling | | |
| | $AB_1$ | 3.784 Gyr |

**Note.** The control model (Model C) has the same initial conditions as Model 2 in Debattista et al. (2020). The left-hand column describes the state of the bar at the time the SMBH is introduced in the model. The middle column gives the model name. The right-hand column gives the time (in gigayears for our choice of model parameters) at which the SMBH is inserted and starts to grow from its initial value of $0.02\ M_{fin}$.

fraction in M31 (Bender et al. 2005; Tamm et al. 2012). Since the SMBH potential is free to move due to accelerations from other particles, we introduce it with an initial mass of $0.02 \cdot M_{fin}$. Through testing we have found this mass (equal to 168 times the disk particle mass) is the minimum necessary to reduce early random motion of the SMBH and is required to ensure the SMBH does not accelerate away from the galaxy center while at ∼0 mass.

To introduce the SMBH into a model at a specific time, we use a snapshot of Model 2 evolved with no SMBH until the desired time as the initial conditions for a new model. Our SMBH potential is then added and evolved as in Equation (5). In this way we may grow an SMBH at a known stage in the evolution of a model by branching a new model off from the case where no SMBH is present.

In all models we keep the final mass ($M_{fin}$), the initial mass, the growth period ($t_{grow}$), and Plummer potential softening length fixed (see values listed in Table 1) and only vary the time of introduction of the SMBH. Previous works have shown that long-term effects on measurable bar quantities are fairly independent of $t_{grow}$, but depend strongly on $M_{fin}$ and softening length (Shen & Sellwood 2004). Since our tests showed similar results, we consider a single growth period of $t_{grow} = 50$ dynamical times in simulation units, equivalent to 378 Myr.

We use an SMBH softening length $\varepsilon = 33.33$ pc, or two-thirds of the global $\varepsilon = 50$ pc (for both disk and halo particles).[8] This is more diffuse than the most compact CMCs meant to represent an SMBH with $\varepsilon \sim$ few parsecs, such as in Shen & Sellwood (2004), but still much more compact than simulated gas concentrations and star clusters (Shen & Sellwood 2004; Athanassoula et al. 2005; Sellwood & Gerhard 2020). This choice was motivated primarily by computational considerations; reducing $\varepsilon$ for the given mass greatly increases the time resolution requirements in the vicinity of the SMBH. We choose a base time resolution of 0.03784 Myr (0.005 dynamic times) for all models. At larger radii, particle motion is calculated every $2^N$ time steps for $N$ zones beginning at 2.4, 7.2, 12, and 19.2 kpc (multiples of scale radii). The criterion of Shen & Sellwood (2004) demonstrates that our chosen softening length is large enough that a circular orbit on the scale of $\varepsilon$ will be sufficiently resolved, i.e., an orbital period will take $\gtrsim 100$ time steps. Increasing base time resolution to the extent required to reduce the SMBH softening length by a factor of 5–10 was prohibitively expensive.[9,10]

Simulation parameters required to evolve these models using GALAXY version 15.4, including full details of grid structure, are included in Appendix B. Simulation snapshots used for analysis were saved every 800 time steps; therefore, the time resolution for results presented in plots is 30.3 Myr unless otherwise noted.

### 2.3. Overview of Models

The results presented in most of this paper are based on a small handful of models that we refer to as the principal set, detailed in Table 2. Each represents a scenario in which an SMBH is introduced into a snapshot of the control model, Model C, which is evolved from the initial conditions of Debattista et al. (2020) Model 2 with no SMBH. All other models in Table 2 are grown from snapshots of Model C with the SMBH introduced at the indicated time, and advanced forward to reach the same duration of total evolution, $T_{final} = 7.568$ Gyr.

In Model $BF_0$, the SMBH is introduced at $t = 0$, before bar formation i.e., before the $m = 2$ amplitude begins to increase from zero (SMBH growth is completed by $t = 0.387$ Gyr). The three models in which the SMBH is introduced before Model C experiences bar buckling are labeled Model $BB_1$, $BB_2$, and $BB_3$. The buckling time is considered to be the time of peak buckling amplitude (defined below).

These three models branch from Model C at $t = 0.575$, 1.150, and 1.877 Gyr, respectively (corresponding to convenient times in internal simulation units). Finally, to compare with previous work (e.g., Hasan et al. 1993; Norman et al. 1996; Shen & Sellwood 2004; Athanassoula et al. 2005; Hozumi & Hernquist 2005; Du et al. 2017) in which the SMBH was grown after the bar had reached a quasi-equilibrium state, we also ran a model with the SMBH introduced in Model $C$ at $t = 3.784$ Gyr, after buckling has occurred (Model $C$ buckles at 2.85 Gyr), which we call Model $AB_1$. Section 3.3 and Appendix A describe additional models that we explored to assess the generality of our results.

---
[8] We note that Debattista et al. (2017, 2020) also used $\varepsilon = 50$ pc for disk particles, but 100 pc for dark matter halo particles. GALAXY uses a single softening length for all components.

[9] Our models were run using GALAXY version 15.4, which included the option of guard radii to approximate increased time resolution in fixed radial regions, which advanced particle motions at smaller $\Delta t$ than the base time step. However, the potential was not recalculated between base time steps. This is justified only for cases when the potential is dominated by a constant value in known radial regions (such as a rigid potential at the origin). However in our case the SMBH moved sufficiently that this justification did not apply at the small scales required. Therefore, we did not implement guard radii. See Shen & Sellwood (2004) for an in-depth description of their use.

[10] The newest versions of GALAXY, version 16.0 and onward, allow the time step at which a particle's coordinates are advanced to be chosen according to the acceleration it is experiencing, which would likely mitigate some of these issues.

## 3. Impact of SMBH Growth on Bar Morphology

To quantify and compare the bar large-scale properties, we employ the commonly used measurements of bar amplitude ($A_{m=2}$) and buckling amplitude ($A_{buck}$; e.g., Sellwood & Athanassoula 1986; Debattista et al. 2020). These quantities are defined using the $m=2$ symmetry mode of an azimuthal Fourier transform of the face-on disk, normalized by the $m=0$ mode as follows:

$$A_{m=2} = \left| \frac{\sum_k m_k e^{2i\phi_k}}{\sum_k m_k} \right|, \quad (6)$$

and

$$A_{buck} = \left| \frac{\sum_k z_k m_k e^{2i\phi_k}}{\sum_k m_k} \right|, \quad (7)$$

where $m_k$, $\phi_k$, and $z_k$ are the mass, azimuth, and vertical position of the $k$th particle, respectively. $A_{buck}$ is a measure of asymmetry of $m=2$ structures about the midplane of the disk, which is sensitive to the bar bending during buckling. Following previous authors (e.g., Debattista et al. 2020), the sums in the above equations are taken over all disk particles, which allows for more direct comparison between different galaxies. Consequently, although both $A_{m=2}$ and $A_{buck}$ are dominated by contributions from a bar, they may also include contributions from other two-fold symmetric structures such as spiral arms. Spiral structures are significant in our models in only the very early stages of evolution as axisymmetry is first broken, but quickly dissipate as the bar forms, leaving the bar as the dominant contributor to the $m=2$ mode.

### 3.1. Bar Strength and Buckling Amplitude

We first examine the effect of SMBH growth on the bar strength (quantified by bar amplitude $A_{m=2}$) both prior to buckling (when the bar is forming and growing) and at late times. Figure 1 shows $A_{m=2}$ (top) and buckling amplitude $A_{buck}$ (bottom) as a function of time. Each curve corresponds to a different model from the principal set in Table 2. Model $C$ (blue) is the control model without an SMBH. The other curves show models with a growing SMBH, and the vertical bands show the period during which the SMBH grows in the model with the curve of the same color. In all models, bar buckling is characterized by a sharp drop in $A_{m=2}$ in the top panel and a corresponding spike in $A_{buck}$ in the lower panel.

The early rise and drop in $A_{m=2}$ prior to $\sim 1.1$ Gyr is associated with formation then dissolution of a strong $m=2$ spiral pattern. Although a weak bar first forms during this time, its contribution to the measured $m=2$ mode is subordinate to the spiral. During this time no buckling occurs (despite the drop in $A_{m=2}$), as is evident from the fact that $A_{buck} \sim 0$. After the drop at $t \sim 1.1$ Gyr, the bar becomes the dominant $m=2$ structure, and remains dominant for the remainder of the time.

Model $C$ (no SMBH) buckles the earliest with the lowest bar amplitude prior to buckling. The lower panel also shows that this model has the greatest buckling amplitude. After buckling, its bar ceases growth and quickly reaches an approximately steady state in bar amplitude that lasts throughout the simulation period. In contrast, Model $BF_0$ (orange), where the SMBH starts to grow at $t=0$, buckles later than all other models with a much lower spike in $A_{buck}$ than Model $C$. Model $BF_0$ has the largest $A_{m=2}$ at late times. This difference in bar amplitude is also apparent shortly after buckling, before any further bar growth occurs.

Figure 1 (top) shows similar trends for the other models with SMBHs introduced before buckling: SMBHs that start growing later (but before bar buckling) have smaller late-time bar amplitudes than Model $BF_0$, but all have higher $A_{m=2}$ than Model $C$. Consistent with previous work (e.g., Shen & Sellwood 2004; Athanassoula et al. 2005; Hozumi & Hernquist 2005), Model $AB_1$ (brown curve) in which the SMBH is grown after bar buckling, when bar growth has ceased, is the only model in which $A_{m=2}$ at late times is lower than in Model $C$.

### 3.2. Weakening and Delay of Buckling

Figure 1 (bottom) shows that in cases in which the SMBH is introduced prior to bar buckling, the buckling event is partially suppressed. We define "suppression" as either the reduction of buckling amplitude ($A_{buck}$), a delay of the buckling event (location of the peak in $A_{buck}$), or both, relative to Model $C$. In Models $BF_0$, $BB_1$, and $BB_3$, buckling is both weakened and delayed. Model $BB_2$ produces a strong buckling event, almost as strong as the buckling experienced by Model $C$; however, it is still significantly delayed compared to Model $C$ such that the final bar amplitude at late times is still significantly larger than in Model $C$. We assert that since buckling weakens the bar, the delay in buckling allows the models to develop a stronger bar prior to buckling. Simultaneously a generally weaker buckling event is associated with a smaller decrease in bar amplitude during buckling. Therefore, both effects contribute independently to a greater final bar strength than in Model $C$. The reasons for this suppression of buckling are explored in depth in Section 4.

Finally we draw attention to the long-term $A_{buck}$ behavior of Model $BF_0$, which after buckling remains elevated relative to other models and gradually increases with time for the duration of the simulation (Figure 1 lower panel). This is indicative of a persistent $m=2$ asymmetry about the midplane (in both the bar and disk). In other words, the bar and disk remain bent throughout the duration of the simulation rather than only bending out of plane for a short period during buckling. This is similar to behavior encountered, but typically not elaborated upon, in multiple other works (e.g., Gardner et al. 2014; Li & Shen 2015; Debattista et al. 2017; Smirnov & Sotnikova 2018), and is poorly understood. Cuomo et al. (2023) carried out a detailed study of such long-term asymmetry about the midplane in both simulations and observations. We do not investigate this asymmetry further in this paper, although we note that similar long-term bending was also observed in a few other simulations with other initial conditions that we simulated (see, e.g., Figures 2 and 3). This long-term bending is seen in simulations with and without SMBH and has also been found by other authors (all without SMBH).

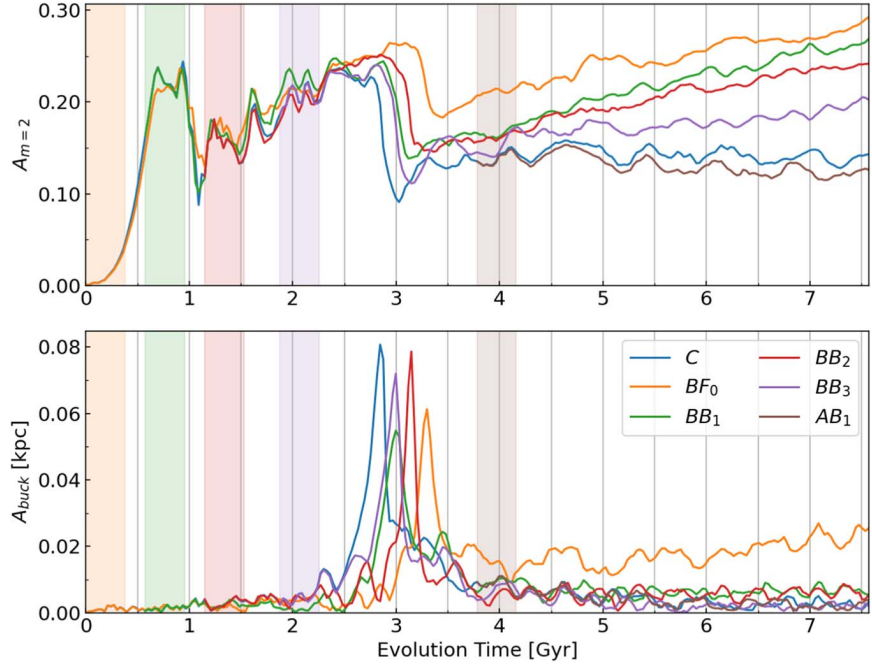

**Figure 1.** Plots of the globally measured bar quantities as a function of time. Top: the bar amplitude ($A_{m=2}$, scaled $m = 2$ Fourier amplitude); Bottom: the buckling amplitude ($A_{buck}$, scaled $m = 2$ asymmetry about the midplane). Each color represents a different model grown as a branch from Model $C$. The curves therefore begin at the time the SMBH is introduced. The SMBH growth period for each model is indicated by the vertical shaded regions of the same color.

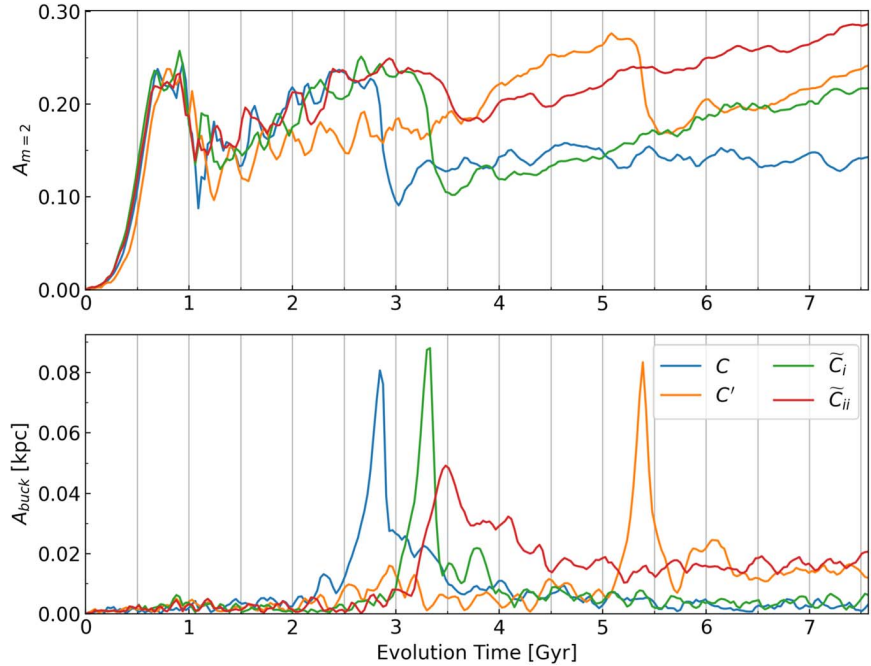

**Figure 2.** Similar to Figure 1, to test the impact of stochasticity on the initial conditions of Model $C$ (see the text for details on how initial conditions for $C'$, $\widetilde{C}_i$, and $\widetilde{C}_{ii}$ were generated). All models were run under identical parameters to Model $C$.

### 3.3. Tests of Stochasticity and Sensitivity to Initial Conditions

Sellwood & Debattista (2009) have shown that large $N$-body systems are inherently chaotic and thus susceptible to stochastic effects, where even small differences in the initial conditions, even for the same model parameters can build to significant differences late in the simulation (affecting quantities such as bar growth rates and susceptibility to repeated buckling). These authors assert that all outcomes are, however, equally valid representations of the evolution of the system. In the previous subsections, Model $C$ was evolved from the same set of initial conditions as Model 2 of Debattista et al. (2020), which are evolved for a longer period by Anderson et al. (2022). An observant reader will notice that while the bar amplitude of Model $C$ does not increase after the buckling event at $t = 2.85$ Gyr, the bar amplitude in Model 2 of Anderson et al. (2022; which first buckles at a similar time $t = 2.9$ Gyr) continues to grow and eventually buckles a second time. We draw attention to this difference as a possible example of stochasticity, where the two models diverge in behavior gradually and relatively late in their evolution (in this case, a small difference in how hot the disk is at the time after buckling). Sellwood & Debattista (2009) demonstrated that such effects can grow from differences as minute as a difference in memory allocation between two processors or a change in the order in which particle coordinates are listed in the initial conditions, with all other factors being equal. Therefore, a late time divergence of model behaviors (even beginning from exactly the same initial conditions) evolved using entirely different simulation codes, such as GALAXY by us and PKDGRAV (Stadel 2001) by Anderson et al. (2022) is virtually unavoidable. This was already demonstrated by Sellwood & Debattista (2009) in a particularly relevant direct comparison of initial conditions evolved using GALAXY and PKDGRAV.

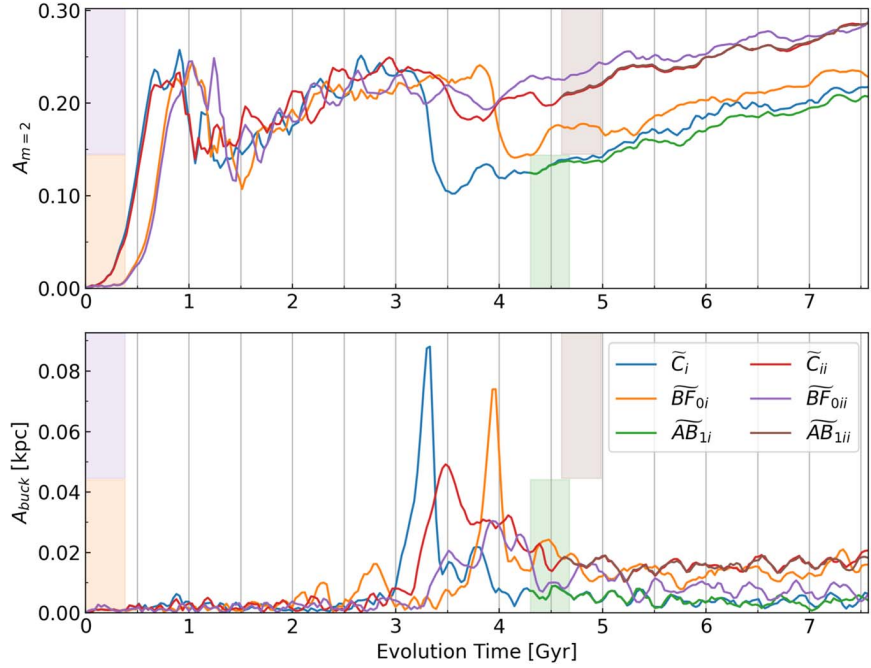

**Figure 3.** Similar to Figure 1. Models $\widetilde{BF}_{0i}$, $\widetilde{BF}_{0ii}$ are analogous to $BF_0$, and Models $\widetilde{AB}_{1i}$, $\widetilde{AB}_{1ii}$ are analogous to $AB_1$ but are evolved from the corresponding scrambled initial conditions $\widetilde{C}_i$, $\widetilde{C}_{ii}$. The vertical orange and purple bands at $\sim t = 0$ show where the SMBH in the $BF_0$ analogs grow and the green and brown bands show where the SMBH in the $AB_1$ analogs grow. Similarly to Figure 1, the early introduction of an SMBH in Models $\widetilde{BF}_{0i}$, $\widetilde{BF}_{0ii}$ does not reduce the bar amplitude in absolute terms.

As a further test for sensitivity to stochastic effects with the same code (GALAXY) and same parameters, we ran two additional tests starting with the initial conditions for Model $C$ presented in the previous two subsections.

1. We use a feature provided by the GALAXY code that allows various sectoral harmonics of the Fourier expansion of the potential to be turned on/off, thereby turning on/off particular forces in calculation and allowing for enforcement of certain symmetries. We disabled all modes other than $m = 0$, and evolved Model $C$ for 48 dynamical times (363 Myr). In this case, we did not grow an SMBH during the time the higher Fourier modes were turned off. This model effectively resulted in a new set of axisymmetric equilibrium initial conditions with overall model parameters identical to Model $C$. We note that the suppression of all modes except $m = 0$ results in a disk with somewhat lower radial velocity dispersion than in Model $C$, but with otherwise identical properties. This simulation was treated as a separate set of initial conditions, and evolved further without an SMBH (Model $C'$) and with an SMBH inserted at $t = 0$ analogous to $BF_0$ (Model $BF_0'$). Since Model $C'$ has smaller radial velocity dispersion than Model $C$, the bar strength initially increases more slowly than in Model $C$ and buckles significantly later than in Model $C$ (see orange curve in Figure 2). As can be seen in Figure 2, the bar in Model $C'$ continues to grow in amplitude after buckling (orange curve) unlike Model $C$ (blue curve). In Model $BF_0'$ (not shown), the final bar strength was stronger than the bar in Model $C'$ at the final time step and is also stronger than for Model $BF_0$.

2. We azimuthally scrambled the initial conditions of Model $C$, creating two more axisymmetric models by generating random azimuthal angles for all particles (and appropriately rotating their Cartesian planar velocities). The two models dubbed $\widetilde{C}_i$ and $\widetilde{C}_{ii}$ were created by the same process, with the only difference being the starting value of the random seed used to scramble (see green and red curves in Figure 2). As can be seen in this figure both these scrambled initial conditions buckle slightly later than Model $C$ (but before Model $C'$). In particular Model $\widetilde{C}_{ii}$ buckles only weakly and ends up with a higher bar amplitude than any of the other Model $C$ analogs.

We also studied the effect of both early-growing black holes ($BF_0$ analogs) and black holes introduced after bar buckling ($AB_1$ analogs) on the final bar strength. Since the scrambled control models ($\widetilde{C}_i$, $\widetilde{C}_{ii}$) do not reach steady state (unlike Model $C$), for the $AB_1$ analogs we introduce the SMBH the same number of dynamical time units after bar buckling as we do for model $AB_1$. Figure 3 shows Model $C$ analogs ($\widetilde{C}_i$, $\widetilde{C}_{ii}$), Model $BF_0$ analogs ($\widetilde{BF}_{0i}$, $\widetilde{BF}_{0ii}$), and Model $AB_1$ analogs ($\widetilde{AB}_{1i}$, $\widetilde{AB}_{1ii}$). To avoid crowding, we do not show models $C'$, $BF_0'$, and $AB_1'$ but the latter two models show a similar degree of stochasticity to the models shown in Figure 3.

Both $\widetilde{BF}_{0i}$ and $\widetilde{BF}_{0ii}$ buckle later and more weakly than their corresponding control models. As in $BF_0$, the bar amplitudes in $\widetilde{BF}_{0i}$ and $\widetilde{BF}_{0ii}$ continue to grow after buckling. Since Model $\widetilde{BF}_{0i}$ buckles later than Model $BF_0$, it does not attain as high an amplitude as $BF_0$ but it still has a stronger bar than its counterpart without an SMBH (Model $\widetilde{C}_i$). In contrast, both Model $\widetilde{C}_{ii}$ (red) and Model $\widetilde{BF}_{0ii}$ (purple) buckle so weakly that they have similar bar strengths at the end of the simulation. However, even in this case, the buckling amplitude is lower for the model with the SMBH, and buckling is delayed relative to the control, confirming our findings in Section 3.2.

The Model $AB_1$ analogs also show interesting results. In the case of $\widetilde{AB}_{1i}$ (green), the bar is slightly weakened relative to the control model ($\widetilde{C}_i$) (blue). However, introduction of the SMBH does not stop bar growth; therefore, it does not weaken or destabilize the bar in absolute terms. In the case of $\widetilde{AB}_{1ii}$ (brown), the final bar strength is identical to that of the control model $\widetilde{C}_{ii}$ (red), which, as we saw in Figure 2, has the strongest bar amplitude of all of the control models.

There is clearly a great deal of stochasticity in the models we have presented in this section. However, we can confidently state that an SMBH of ≲0.2% of disk mass that grows prior to bar formation never destabilizes or weakens a bar. The theoretical analysis presented in Sections 4 and 5 sheds light on the empirical results found in this section. Although an SMBH introduced after buckling might slightly weaken a bar relative to a case with no SMBH, it does not always do so, especially if the bar is still growing (as we see for Model $\widetilde{AB}_{1ii}$).

Since the previous generation of simulations all introduced an SMBH after the bar had attained a steady state (e.g., Hasan et al. 1993; Norman et al. 1996; Shen & Sellwood 2004; Athanassoula et al. 2005; Hozumi & Hernquist 2005; Du et al. 2017) and since our principal control Model $C$ is the only one that attains a steady state after buckling, in the remainder of this paper we will primarily focus on Model $C$ and the models derived from it that are listed in Table 2.

### 3.4. Strength of the Boxy-peanut/X-shaped Bulge

We now examine how the growth of an SMBH affects the strength of the BP/X-shaped bulges that form in our simulations. Figure 4 shows $x$–$z$ projections of the surface density of final snapshots of four models. The BP/X-shaped bulge is evident, and the models with stronger bars (Models $BF_0$, $BB_1$) clearly show stronger BP/X shapes than models with weaker bars (Models $C$, $AB_1$).

We parameterize the shape of the BP/X bulges using a new bar-deprojection method developed by Dattathri et al. (2023), which we briefly describe here. We construct parametric 3D bar models that are added as additional modules to the surface-brightness fitting routine IMFIT (Erwin 2015). IMFIT is able to take any input 3D distribution and project it at an arbitrary orientation to derive the best-fit parameters of the 3D model required to fit the image. In the new BP/X bar fitting module the bar is assumed to have a sech$^2$ profile in a scaled radius given by

$$\rho = \rho_0 \text{sech}^2(-R_s), \quad (8)$$

where

$$R_s = \left(\left[\left(\frac{x}{X_{\text{bar}}}\right)^{c_\perp} + \left(\frac{y}{Y_{\text{bar}}}\right)^{c_\perp}\right]^{c_\parallel/c_\perp} + \left(\frac{z}{Z_{\text{bar}}}\right)^{c_\parallel}\right)^{1/c_\parallel}. \quad (9)$$

The origin of the coordinate system is the galaxy center; $X_{\text{bar}}$, $Z_{\text{bar}}$, and $Y_{\text{bar}}$ are the semiaxis lengths of the bar along the long-axis ($x$), the axis perpendicular to the disk plane ($z$), and $y$ is the other axis in the disk plane (these describe an ellipsoidal bar). $c_\parallel$ and $c_\perp$ control the diskiness/boxiness of the bar (the 3D analog of the parameters proposed by Athanassoula et al. 1990). While Dattathri et al. (2023) explored a range of values for $c_\parallel$ and $c_\perp$, here we set $c_\parallel = c_\perp = 2$ (corresponding to an ellipsoidal bar). The BP/X shape is determined by a scale height perpendicular to the disk that depends on the position in the $x-y$ plane represented by a double Gaussian centered at the galactic center:

$$Z_{\text{bar}}(x, y) = A_{\text{pea}} \exp\left(-\frac{(x - R_{\text{pea}})^2}{2\sigma_{\text{pea}}^2} - \frac{y^2}{2\sigma_{\text{pea}}^2}\right)$$
$$+ A_{\text{pea}} \exp\left(-\frac{(x + R_{\text{pea}})^2}{2\sigma_{\text{pea}}^2} - \frac{y^2}{2\sigma_{\text{pea}}^2}\right) + z_0, \quad (10)$$

where $R_{\text{pea}}$ and $\sigma_{\text{pea}}$ are the distance of the center of the peanuts from the galaxy center and the width of each peanut, respectively. $A_{\text{pea}}$ measures the vertical extent of the peanut above the ellipsoidal bar $z_0$ (not the distance from the midplane), since $z_0$ is the base vertical scale height of the bar. The maximum distance of the peanut from the midplane is therefore given by

$$h_{\text{pea}} \approx z_0 + A_{\text{pea}}, \quad (11)$$

which we refer to as the "peanut height," which is the maximum value of $Z_{\text{bar}}$. We note that in our model the peanut is entirely associated with the bar, and there is no separate bulge component.

Equation (10) is similar to the "peanut height function" proposed previously by Fragkoudi et al. (2015), with the additional assumption that both halves of the peanut are symmetric about the origin, and the peanuts are aligned with the major axis of the bar.

The shape of the bar is controlled primarily by three parameters: $R_{\text{pea}}$, $h_{\text{pea}}$, and $\sigma_{\text{pea}}$. Dattathri et al. (2023) demonstrated that these three parameters offer great versatility for modeling a variety of shapes for the BP/X feature. They also show, by fitting mock 2D surface-brightness data generated from some of the $N$-body models in this work, that the 3D density distribution and potential arising from this parameterization of the bar and BP/X bulge closely match the $N$-body density and potentials. Dattathri et al. (2023) also showed that the BP/X shape parameters can also be obtained for an $N$-body model by fitting Equation (10) directly to the stellar particle distribution in order to obtain the "true" values of $R_{\text{pea}}$, $h_{\text{pea}}$, and $\sigma_{\text{pea}}$ instead of going through the deprojection algorithm. Dattathri et al. (2023) found that this 3D fit to the

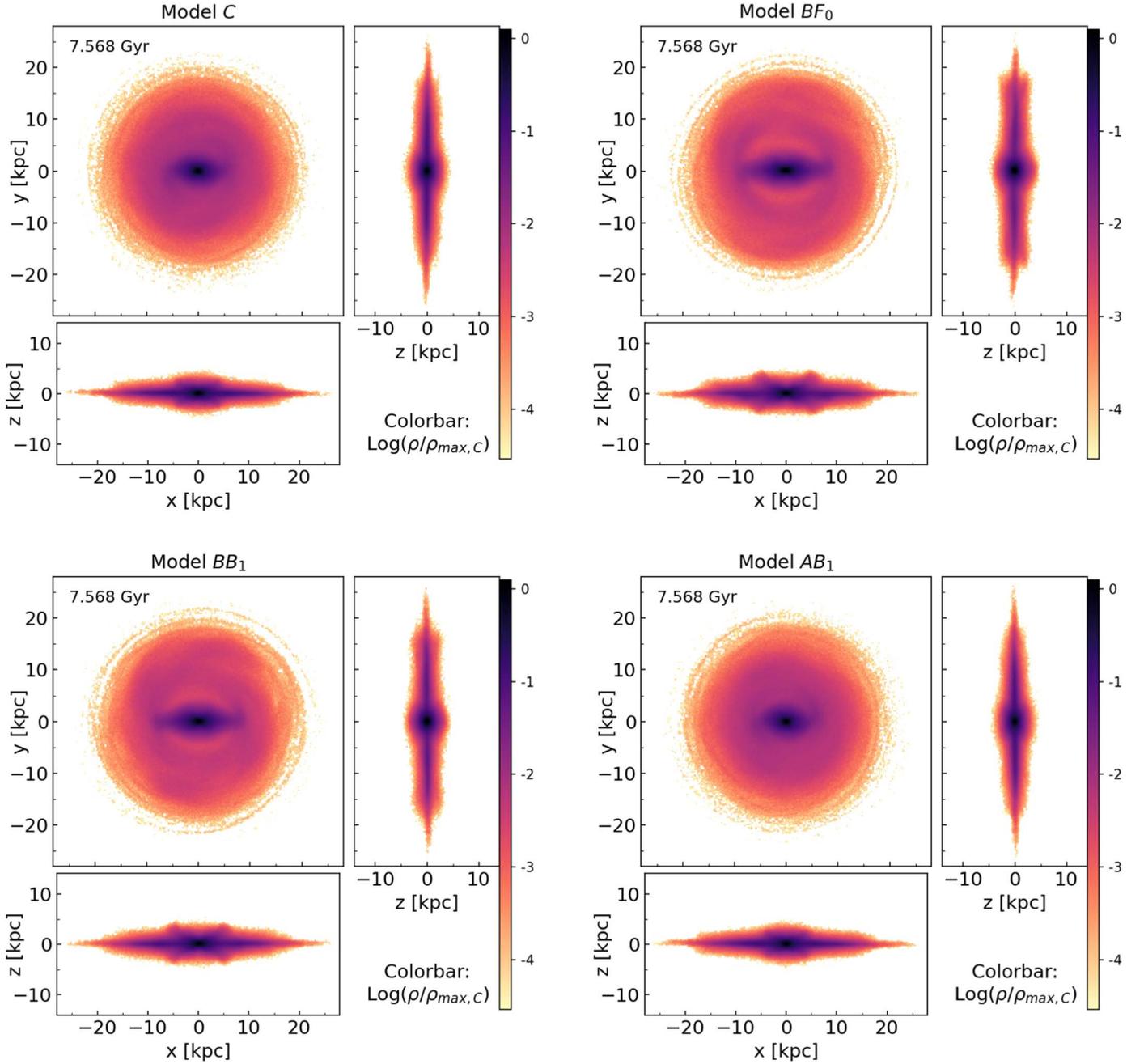

**Figure 4.** Projected logarithmic surface density maps of selected models at the final snapshot of the simulation (evolution time = 7.568 Gyr). Each projection orientation is normalized by the maximum projected surface density of Model *C* viewed from that orientation. Therefore, for each orientation all models share a common normalization in which 0 represents the maximum density of Model *C* in that orientation, and are directly comparable. The increased prominence of the bar and BP/X bulges of models in which an SMBH is grown early (Models $BF_0$, $BB_1$) is clearly evident.

*N*-body particle distribution gives a better representation of the gravitational potential.

Here we use this direct 3D fitting to the *N*-body stellar particle distribution to quantify the strength of the BP/X bulge at the final time step of 12 different models. The key parameters that determine the BP/X shape: $R_{\rm pea}$, $h_{\rm pea}$, and $\sigma_{\rm pea}$, are recovered for each model at the final time step.

Figure 5 shows the values of the parameters $R_{\rm pea}$ and $h_{\rm pea}$ obtained via 3D fitting as a function of the bar amplitude of the model ($\sigma_{\rm pea}$ shows no correlation with bar amplitude and hence is not shown). It is clear that the strength of the peanuts as parameterized by $h_{\rm pea}$ and $R_{\rm pea}$ are strongly correlated with bar amplitude. This strong correlation between the bar amplitude and the parameters ($R_{\rm pea}$ and $h_{\rm pea}$) characterizing the BP/X shape does not depend on whether or not a model includes a black hole (the points with black edges are control models while the others have an SMBH). The cause of this correlation is investigated in Section 6. While further simulations are needed to confirm this correlation in hydrodynamical simulations, the existence of these correlations has an important implication. For edge-on disks in which the $m = 2$ bar amplitude is impossible to measure by traditional means, the use of the BP/X fitting function in Equation (10) could enable a determination of bar strength (if the bar major axis is 25°–90°

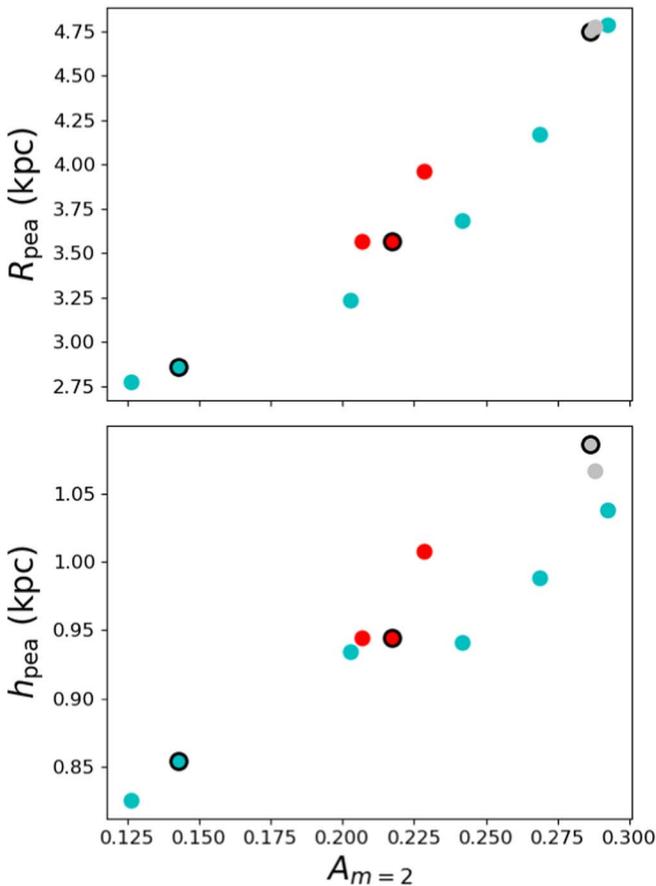

**Figure 5.** The values of $R_{\rm pea}$ (top) and $h_{\rm pea}$ (bottom) plotted as a function of the bar amplitude at the final time step. The points with black edges are the control models. Cyan points are Models $C$, $BF_0$, $BB_1$...; red points are Models $\widetilde{C}_i$ and its $BF_0$, $AB_1$ analogs; and gray points are $\widetilde{C}_{ii}$ and its $BF_0$, $AB_1$ analogs. The gray points overlap significantly since all of these models have nearly identical behavior.

to the line of sight). This will be investigated further in the future.

## 4. Causes of the Weakening/Delay of Bar Buckling

As briefly mentioned in Section 1, several previous papers studying the origin of the buckling instability identified a large velocity anistotropy as one of the primary causes of out-of-plane bending, with in-plane velocity dispersion, $\sigma_R$, significantly higher than the velocity dispersion perpendicular to the plane, $\sigma_z$. Toomre (1966) showed analytically for an idealized, infinite thin sheet that there is a critical value of $\sigma_z/\sigma_R > 0.3$, which when satisfied stabilizes the sheet to out-of-plane bending. This value was subsequently revised by simulations of a finite thin sheet (Araki 1985) and analytic stability analysis for a stellar distribution of finite thickness and finite extent (Merritt & Sellwood 1994), both of which showed a larger critical value of $\sigma_z/\sigma_R > 0.6$ where the systems were stable to bending. The higher value of $\sigma_z/\sigma_R > 0.6$ was confirmed for simulated triaxial ellipsoids and finite disk systems by Sellwood & Merritt (1994). While the precise critical value of $\sigma_z/\sigma_R$ depends on the complexity of the system, it is clear that an increase in $\sigma_z/\sigma_R$ in the disk decreases the susceptibility of the system to out-of-plane bending/buckling. The buckling event itself significantly redistributes the stellar kinetic energy from the radial direction to the vertical direction causing a sharp and sudden increase in $\sigma_z/\sigma_R$, and therefore generally stabilizing the system against subsequent bending.

Since previous work shows that the value of $\sigma_z/\sigma_R$ determines the susceptibility to buckling, we now compute this quantity for our simulations. In this section we focus on our principal suite of simulations Model $C$, $BF_0$, $BB_1$, $AB_1$ although we briefly discuss this anisotropy for one of the other sets of models ($\widetilde{C}_{ii}$ and its derivatives) in Appendix A. For each snapshot, we compute both $\sigma_R$ and $\sigma_z$ for disk particles in 257 radial bins with $\sim 2.3 \times 10^5$ star particles in each bin. We disregard the outermost bin containing the most diffuse particle distribution on the outskirts of the disk, leaving 256 bins with a maximum extent of $R = 19$ kpc. Figure 6 shows contour plots of $\sigma_R$ (left), $\sigma_z$ (middle), and $\sigma_z/\sigma_R$ (right), each as a function of time (x-axis) and radius $R$ (y-axis) for the four models ($C$, $BF_0$, $BB_1$, $AB_1$).

In the top row, for Model $C$ (similar to other models at early times), we see that the formation of the bar ($t \lesssim 1$ Gyr) coincides with an increase in $\sigma_R$ for $R \lesssim 4$ kpc (left columns), while $\sigma_z$ is much less significantly changed. This leads to a decrease of $\sigma_z/\sigma_R$ through much of the disk (manifesting as a deep blue hue at $0.5 < t/{\rm Gyr} < 2.2$). As the bar grows, $\sigma_R$ continues to grow in magnitude in the inner disk, while $\sigma_z$ also increases at a rate such that $\sigma_z/\sigma_R$ remains at a relatively constant value in the inner disk until the bar buckles—see also Figure 1 (bottom panel). During bar buckling (peak buckling time denoted by vertical white line), the bending and thickening of the bar results in significant vertical heating, thus rapidly increasing $\sigma_z$ over a short time. Buckling also radially shortens the bar on a similar timescale, resulting in decreased $\sigma_R$. This produces a very rapid increase of $\sigma_z/\sigma_R$ over much of the inner disk. All radii within $\sim 5$ kpc reach values of $\sigma_z/\sigma_R \gtrsim 0.6$ and appear to become stable to buckling.

In the second and third rows of Figure 6, we see that the introduction of an SMBH prior to buckling very quickly begins to produce more vertical heating in the center of the disk upon introduction, due to scattering of stars by the SMBH. This increases $\sigma_z$ significantly, thus increasing $\sigma_z/\sigma_R$. This can be seen as the appearance and widening in radius of red-colored regions in the right-hand contour plots (i.e., $0.6 < \sigma_z/\sigma_R < 0.7$) at small radii soon after the white-shaded vertical region marking the growth of the SMBH (i.e., as early as 0.5 Gyr in model $BF_0$ and by 1 Gyr in model $BB_1$). This increase in $\sigma_z/\sigma_R$ slowly propagates to larger radii ($\sim 1$ kpc), well beyond the expected SMBH sphere of influence, $r_{\rm BH} = 0.1$–$0.175$ kpc[11] when the SMBH has reached full mass. $r_{\rm BH}$ is greatest for Model $BF_0$, and less for other models. This increase to $\sigma_z/\sigma_R \gtrsim 0.6$ in the inner region appears to be responsible for reducing the susceptibility of the bar to buckling. However, this effect is fairly localized and does not propagate outward rapidly enough to completely suppress buckling at all radii. Nonetheless, as demonstrated in Figure 1, it clearly delays and/or weakens buckling in all of our models with an early-growing SMBH.

In the fourth row of Figure 6, we see that the introduction of the SMBH after buckling (Model $AB_1$) has very little to no effect on $\sigma_z$, and thus has no noticeable effect on long-term $\sigma_z/\sigma_R$ when compared to Model $C$. This is because the bar is already significantly vertically heated during buckling.

---

[11] Here we use the radius from the center within which the mass of star+dark matter particles equals the mass of the SMBH, a quantity that varies slightly depending on the models.

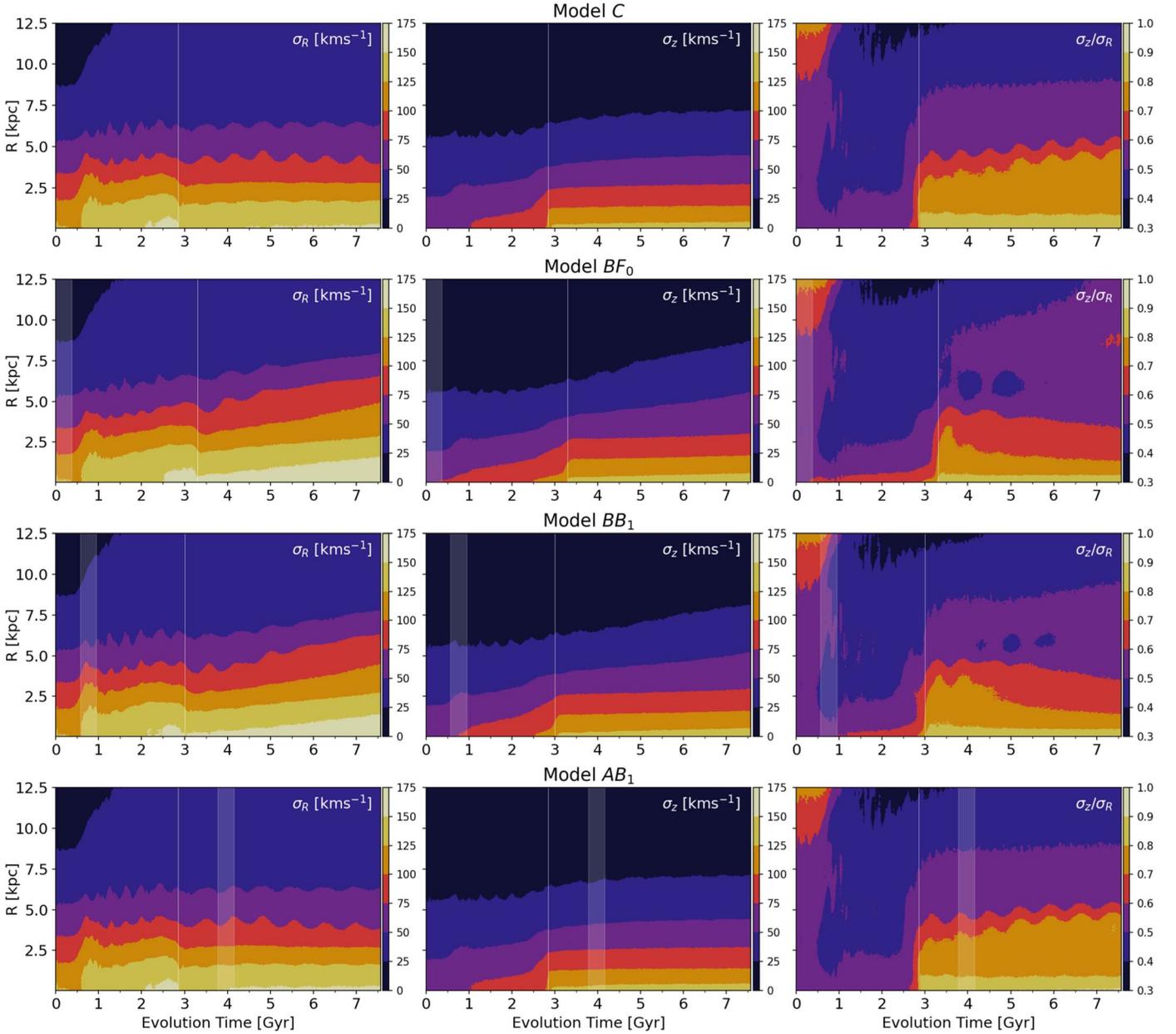

**Figure 6.** Evolution of velocity dispersion and velocity anisotropy as a function of radius in the stellar disk. Each row corresponds to a particular model as labeled. The left column shows radial velocity dispersion ($\sigma_R$), the center column shows vertical velocity dispersion ($\sigma_z$), and the right column shows the ratio ($\sigma_z/\sigma_R$). In cases where an SMBH is present, a transparent shaded region marks the growth period of the SMBH beginning at its introduction, ending when the SMBH has reached full mass. The time of peak buckling amplitude for each model is indicated by a thin vertical white line. $\sigma_z/\sigma_R$ begins to increase shortly after introduction of the SMBH in Models $BF_0$ and $BB_1$ thereby partially stabilizing the disk against buckling, whereas in Model $C$, $\sigma_z/\sigma_R$ remains effectively constant at all radii until buckling.

We note that while $\sigma_R$ does not appear to change much with the introduction of the SMBH, in Models $BF_0$ and $BB_1$, $\sigma_R$ is significantly higher than in Models $C$ and $AB_1$ (note the very light yellow color indicating $\sigma_R \gtrsim 150\,\mathrm{km\,s^{-1}}$). We will return to discuss the significance of this difference in Section 6.2.

When models $BF_0$ and $BB_1$ do buckle, the reduced magnitude of buckling can also be observed from Figure 6. The change in $\sigma_z/\sigma_R$ is not as large at most radii as in the control Model $C$, which is primarily due to a comparatively less significant decrease in $\sigma_R$, despite a similar or greater increase in $\sigma_z$ due to SMBH effects prior to buckling. This implies that during the strong buckling of Model $C$ the value of $\sigma_z/\sigma_R$ in the inner disk crests well above the stability threshold value.

While the bar in Model $C$ does not grow after buckling, the continued growth of the bar amplitude after buckling observed in Models $BF_0$, $BB_1$, $BB_2$, and $BB_3$ (in Figure 1 top) manifests as an increase of $\sigma_R$ throughout the disk. $\sigma_z$ also continues to gradually grow in the inner disk, a consequence of the increase in bar strength, but in the models with an SMBH, $\sigma_z/\sigma_R$ actually decreases due to the increase in $\sigma_R$ over a range of radii compared to Model $C$, where $\sigma_z/\sigma_R$ remains effectively constant after buckling (since the bar does not continue to grow). Despite the slight decrease in $\sigma_z/\sigma_R$ in the models with an SMBH, $\sigma_z/\sigma_R$ remains above 0.6 for $R < 5$ kpc. This is noteworthy since it explains why the models with an SMBH are not prone to a second buckling episode despite the steadily increasing bar amplitude (on the range of timescales we explored). We evolved Model $BF_0$ for an additional 7.5 Gyr (15 Gyr in total), but found no evidence of further buckling.

To summarize the results of this section, we conclude that the primary reason for the early-growing SMBH to cause both an increase in bar amplitude and either a delay or a weakening of bar buckling (or both) is that the SMBH causes rapid vertical heating of the stellar disk almost as soon as it is introduced. The extent of the vertical heating increases in radius with time, greatly exceeding the nominal sphere of influence of the black hole, in large part because the stars in the bar have highly radial orbits and therefore can experience scattering by the central black hole even if they have large apocenter radii. Since these bar orbits are already highly radial, their radial velocity dispersion is not significantly affected, and the primary effect is an increase in $\sigma_z/\sigma_R$ to values about ∼0.6 or greater, increasing the stability of the disk. This delays bar buckling and/or reduces the buckling amplitude.

## 5. The Role of Resonances and Resonant Orbits in Bar and BP/X Strengthening

We show in Figures 4 and 5 that simulations with stronger bars also have bulges with stronger BP/X shapes. The BP/X shape has been attributed to either bar buckling (Raha et al. 1991; Merritt & Sellwood 1994; Sellwood & Merritt 1994; Debattista et al. 2017; Collier 2020) or resonant trapping of orbits into the vertical inner Lindblad resonance (ILR; Combes et al. 1990; Pfenniger & Friedli 1991; Quillen 2002; Quillen et al. 2014; Sellwood & Gerhard 2020). It is clear from the previous section that the buckling instability is weakened by the presence of an early grown SMBH (which causes vertical disk heating) while, paradoxically, the BP/X shape is strengthened. We therefore explore whether the role of resonances and resonant trapping by the ILR and vILR is enhanced by an SMBH.

In rotating disks, resonances arise between the vertical oscillation frequency $\nu(R)$, the epicyclic (radial) oscillation frequency $\kappa(R)$, the circular orbit frequency $\Omega(R)$, and the bar pattern speed $\Omega_p$. The ILR occurs when $\Omega_p = \Omega(R) - \kappa(R)/2$. Likewise, the vertical inner Lindblad resonance (vILR) arises when $\Omega_p = \Omega(R) - \nu(R)/2$. Strictly speaking, the definitions above only apply to axisymmetric or weakly nonaxisymmetric systems, but we use them in Section 5.1 for a qualitative discussion. Although we do not show Models $BB_1$, $BB_2$, and $BB_3$ to avoid crowding, all models in which the SMBH is introduced prior to buckling develop trends similar to model $BF_0$, soon after the SMBH is introduced.

In Section 5.2 we use the powerful framework of orbital frequency analysis and apply it to the orbits of $10^5$ stars selected in multiple snapshots of Models $C$, $BF_0$, and $AB_1$ in order to identify the most important resonances in the nonaxisymmetric potentials. (We do not show other models, but the orbital frequencies of the $BB$ models contain features intermediate between Model $C$ and Model $BF_0$.) Performing a frequency analysis in both cylindrical and Cartesian coordinates, we arrive at an explanation for why bars can be strengthened in the presence of an early-growing black hole, while an SMBH that has grown after the bar reaches a steady state weakens the bar.

### 5.1. Resonances in the Quasiaxisymmetric Approximation

We use the AGAMA package (Vasiliev 2019) to estimate the frequencies assuming a triaxial potential for the star particles and an axisymmetric potential for the halo. Then, at each radius we compute the average between the frequencies along the major and minor axes of the bar.

In order to estimate the approximate location of the vILR and ILR, we compute the pattern speed of the bar at various times. This is done by fitting a straight line to the time evolution of the phase angle of the bar (assumed to increase monotonically with time) for 10 snapshots around the snapshot of interest. The obtained pattern speeds are shown as horizontal dotted lines in each panel of Figure 7 (with lengths corresponding to the bar lengths in that snapshot). We compute the approximate location of the vILR, ILR, and corotation resonances as the radii where the curves for $\Omega(R)$ (thin solid), $\Omega(R) - \kappa(R)/2$ (thick solid), and $\Omega(R) - \nu(R)/2$ (dashed–dotted) intersect the average bar pattern speed $\Omega_p$ around that snapshot—see Figure 7.

At $t = 0$ all of the models are identical, so only Model $C$ is shown. The next four panels show the three simulations at additional snapshots. The bottom-right panel shows the approximate location of the ILR and vILR as a function of time in each of the three models.

At $t = 1.51$ Gyr, in Model $BF_0$ (orange), $\Omega$ rises sharply at small radii (a consequence of the early-growing SMBH and an increase in the stellar density around it) and both $\Omega(R) - \kappa(R)/2$ and $\Omega(R) - \nu(R)/2$ have already begun to increase relative to Model $C$. At $t = 2.27$ Gyr, both the ILR and vILR are present at small radii in the model with an early SMBH but not in Model $C$. This is because of the combined effect of the increase in $\Omega$ and because the SMBH has already started heating the disk vertically, increasing $\sigma_z$, and therefore decreasing $\nu$ enough to cause $\Omega - \nu/2$ to increase. These two factors cause the appearance of both the ILR and vILR at small radii, nearly 0.5 Gyr prior to the appearance of these two resonances in Model $C$ (see bottom-right panel). In model $BF_0$, the ILR and vILR appear at $R \lesssim 1$ kpc by around $t \simeq 2$ Gyr, and their radial location moves rapidly outward between first appearance and the final snapshots (where it is beyond 5 kpc). In Model $C$ (blue), these resonances appear later ($t = 2.5$–3.0 Gyr), and these resonances stay within $R \sim 3.5$ kpc. Model $AB_1$ is only shown after the SMBH finishes growing (after $t = 4.232$ Gyr). In this model, both ILR and vILR have $R < 3.5$ kpc throughout the evolution, similar to Model $C$. The resonance locations in both models $C$ and $AB_1$ show small oscillations around a constant value after ∼3.5 Gyr, which we verified as being due to small oscillations in the estimated pattern speeds.

We note that at $t = 1.51$ Gyr, the vILR is not present in any model. By $t = 4.54$ Gyr, all models have buckled and the vILR is present in all models thereafter, but it is clear that the SMBH, despite having a small nominal sphere of influence ($r_{BH} \lesssim 0.175$ kpc), is able to cause vertical heating in the bar for stars that travel to much larger radii due to the high radial anisotropy of the bar. The bottom-right panel suggests that the vILR does not appear until the bar buckles in Model $C$, but it appears far prior to bar buckling in Model $BF_0$. Although not shown, we note that in Model $BB_1$ too, the vILR appears before the bar buckles, although it appears slightly later than in Model $BF_0$, as would be expected for $BB$ models in relation to $BF_0$ from our previous results.

These results suggest that, since the vILR and ILR in Model $BF_0$ sweep over large radial range, they can resonantly capture a significantly larger fraction of disk stars than they do in Models $C$ and $AB_1$, where they sweep a much smaller radial range.

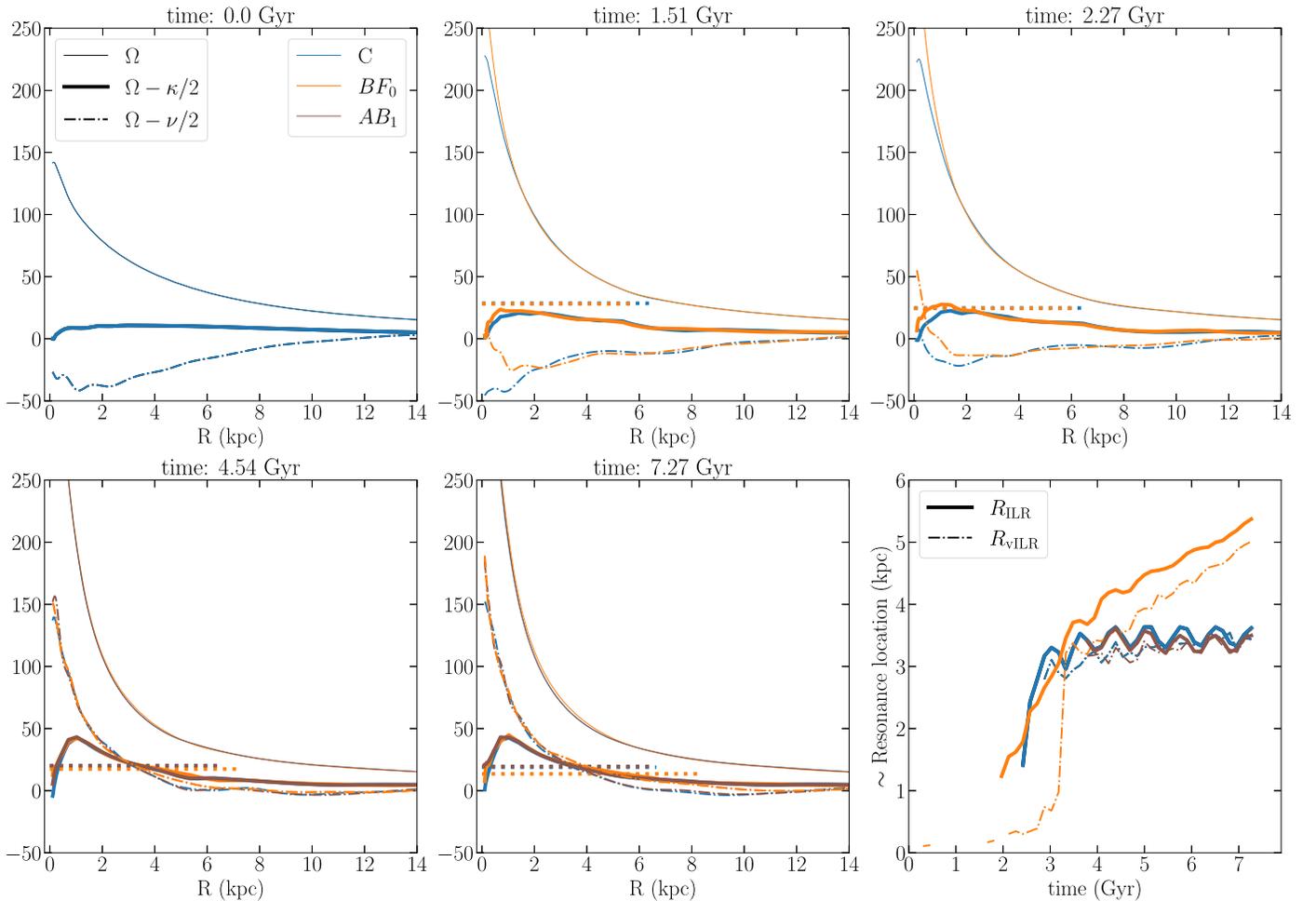

**Figure 7.** Approximate locations of the ILR and vILR in three models (legend) at five times (titles). Horizontal dotted lines show the corresponding bar pattern speeds, with lengths representing bar lengths. Thin solid curves show the rotation frequency $\Omega(R)$, thick solid curves show $\Omega(R) - \kappa/2$, and dashed–dotted curves show $\Omega(R) - \nu(R)/2$ (see the text). The bottom-right panel shows the approximate radius of the ILR (thick solid lines) and vILR (dashed–dotted lines) in each model as a function of time. The model with early SMBH introduction, $BF_0$, develops an ILR and a vILR earlier than the control Model $C$, and these resonances sweep through the disk to much larger radii (~4.5–5 kpc, compared to ~3–3.5 kpc).

### 5.2. Brief Overview of Orbital Frequency Analysis

We now briefly describe the orbital frequency analysis first introduced by Binney & Spergel (1982, 1984). The method was significantly improved and refined by Laskar (1990, 1993) who referred to their algorithm as "Numerical Analysis of Fundamental Frequencies (NAFF)."

In Hamiltonian dynamics, the angle variables and their canonically conjugate actions $J_i$ uniquely define a regular orbit (Binney & Tremaine 2008). In a 3D potential, the $J_i$, $i = 1,..,3$ are conserved, and time derivatives of the angle variables $\Omega_i = \dot{\theta}_i(t)$, $i = 1,..,3$ are also constants of motion. Since regular orbits in galaxies are quasiperiodic, their space and velocity coordinates can be represented by time series of the form:

$$x(t) = \sum_{k=1}^{k_{max}} A_k e^{i\omega_k t} \qquad (12)$$

with similar expressions for $y(t)$, $z(t)$ and velocity components, $V_x(t)$, $V_y(t)$, and $V_z(t)$. The amplitudes $A_k$ of the $N$ (typically $N = 10$–15) largest peaks in the spectrum and their corresponding frequencies $\omega_k$ are obtained by taking a Fourier transform of a complex time series, e.g., $f_x(t) = x(t) + iV_x(t)$ (and similarly for $f_y$, $f_z$) constructed from the spatial and velocity coordinates of an orbit. This is followed by Gram–Schmidt orthogonalization to properly extract additional, lower-amplitude frequencies in the spectrum. Eventually, we obtain the basis set of three linearly independent frequencies $\Omega_i$. All of the other frequencies $\omega_k$ in the three frequency spectra (one each for $f_x$, $f_y$, and $f_z$) can be written as integer linear combinations of these three frequencies; therefore, the $\Omega_i$, $i = 1,..,3$ are referred to as "fundamental frequencies."

Here we use our implementation[12] of the NAFF algorithm to recover orbital fundamental frequencies (Valluri & Merritt 1998; Valluri et al. 2010) and to classify bar orbits (as described in Valluri et al. 2016). With this code, the frequency components in the spectrum can be recovered with high accuracy (1 part in $10^5$ or better) in ~20–30 orbital periods.

Previous work has shown (Valluri et al. 2010, 2013, 2016) that when frequency analysis is applied to a large representative sample of orbits drawn from a self-consistent distribution function, the analysis of the frequency maps (plots showing ratios of orbital fundamental frequencies plotted against each other) and automated orbit classification enables a quantitative assessment of the relative importance of different resonances and different types of orbits to the phase-space structure of the galaxy. It also enables one to understand how the orbital distribution function is altered by evolution of the galactic potential.

---
[12] Publicly available at https://sites.lsa.umich.edu/mvalluri/software/.

Frequency analysis is also a powerful means to assess whether orbits are regular or chaotic. Since regular orbits conserve actions, their frequencies also remain constant in a static potential. By computing each of the three fundamental frequencies $\Omega_i$ over two separate time intervals $T_1$, $T_2$ one can compute

$$\Delta_{\Omega_i} = \left| \frac{\Omega_i(T_1) - \Omega_i(T_2)}{\Omega_i(T_1)} \right| \quad (13)$$

$$\Delta_\Omega = \max(\Delta_{\Omega_i}), \; i = 1,\ldots,3 \quad (14)$$

with $\Delta_\Omega$ defined as the frequency drift. Therefore, frequency drift provides a measure of how chaotic (irregular) an orbit is.

For the analysis of the N-body simulations studied in this paper, we randomly select $10^5$ star particles from Model C in the initial snapshot ($t = 0$). We integrate the orbits of this same set of particles in all of the models at $t = 1.51$, 2.27, 3.03, 3.78, 4.54, and 7.57 Gyr (snapshots 200, 300, 400, 500, 600, and 1000, where the integers denote the number of dynamical times elapsed since the start of the simulation) in the respective potentials and in the presence of the rotating bars. The potentials are estimated with AGAMA, assuming a triaxial distribution for star particles, an axisymmetric distribution for DM particles and a Plummer potential for the SMBH (when one is present). While not all of these star particles eventually end up in the bar in the final snapshot of any specific model, we consider this selection to be sufficiently random and uniform to enable us to compare the results of orbital frequency analysis of the various models in an unbiased manner.

### 5.3. Frequency Analysis in Cylindrical Coordinates: Capture into ILR and vILR

We first compute the orbital frequencies in cylindrical coordinates in the inertial frame of the galaxy, which allows us to properly identify the corotation resonance (Beraldo e Silva et al. 2023). We use the standard cylindrical polar coordinates $R, v_R, \phi, v_\phi, z, v_z$. $R, v_R$, and $z, v_z$ are the canonically conjugate radial and vertical coordinates and momenta, respectively, and hence can be used to define a complex time series $f_R(t) = R(t) + iv_R(t)$ and $f_z(t) = z(t) + iv_z(t)$. However, since $\phi$ is an angular coordinate and $v_\phi$ is an angular velocity (not linear coordinate and linear momentum), this pair does not yield the correct tangential frequencies when used to construct the time series for frequency analysis. Instead we use the angular momentum $L_z = xv_y - yv_x$ and the Poincaré symplectic polar variables $\sqrt{2|L_z|} \cos\phi$ and $\sqrt{2|L_z|} \sin\phi$ to define the complex time series $f_\phi(t) = \sqrt{2|L_z|}(\cos\phi + i\sin\phi)$ (Papaphilippou & Laskar 1996; Valluri et al. 2012). Beraldo e Silva et al. (2023) further discussed the advantages of these complex combinations.

Figure 8 shows histograms of $(\Omega_\phi - \Omega_P)/\Omega_R$ (left) and $(\Omega_\phi - \Omega_P)/\Omega_z$ (right) at different times, color coded by the mean $z_{max}$ in each bin (with width 0.0025)—note the different y-axis scales. The two left columns refer to Model C. The main resonances are easily recognized from the sharp peaks: the corotation ($(\Omega_\phi - \Omega_P) = 0$), the ILR ($(\Omega_\phi - \Omega_P)/\Omega_R = 0.5$; see Athanassoula 2003), and the vILR ($(\Omega_\phi - \Omega_P)/\Omega_z = 0.5$).

The ILR is known as the main resonance supporting bars (e.g., Contopoulos & Papayannopoulos 1980; Athanassoula 2003). The early significant development of this resonance in Model C (upper-left panel of Figure 8) agrees with the early development of the bar observed in Figure 1. It is interesting to note that this early development of the ILR in Model C was not detected in the simpler quasiaxisymmetric approximation—see Figure 7.

In the second column, we see that in Model C the vILR is almost unpopulated at 2.27 Gyr, but a significant number of orbits have crossed this resonance by 4.54 Gyr, i.e., have $(\Omega_\phi - \Omega_P)/\Omega_z \gtrsim 0.5$. Interestingly, the mean $z_{max}$ increases abruptly for these orbits, in agreement with the theoretical expectation of the excitement of vertical motion by this resonance (Binney 1981; Pfenniger & Friedli 1991)—see also Beraldo e Silva et al. (2023).

The two right columns of Figure 8 show the histograms for the model $BF_0$. The ILR (left panels) is similarly populated over time, but it is more populated than Model C already at 1.51 Gyr, and the peak reaches significantly larger values at $t = 4.54$ Gyr. At the final snapshot (bottom row), we identify 54,934 orbits within $\Delta|(\Omega_\phi - \Omega_P)/\Omega_R| \leqslant 0.01$ from the ILR, in comparison to 28,482 identified in the same range for Model C, which agrees with the higher bar amplitude observed for model $BF_0$—see Figure 1.

The right panels show how the vILR is populated over time, as it is already detected in Model $BF_0$ at 2.27 Gyr, and that the mean $z_{max}$ increases abruptly for orbits crossing this resonance (i.e., moving rightward in the histogram). In particular, we note that this resonance is very strongly populated at $t = 4.54$ Gyr and at $t = 7.57$ Gyr (much stronger than in Model C) and that at the final snapshot the mean $z_{max}$ for orbits close to the resonance is substantially larger than in Model C, although the number of stars strictly at the resonance is smaller. The number of orbits that crossed the vILR is also significantly larger than in Model C. Taking into account that the BP/X bulge is stronger in this simulation, this suggests that after stars are released from (or cross) the vILR, they continue to support the BP/X. This seems to agree with the theoretical scenario described by Quillen et al. (2014), where stars do not stay trapped by the vILR for a long time, but crossing this resonance is enough to promote stars to orbits with high $z_{max}$ supporting the BP/X shape—see also Sellwood & Gerhard (2020).

Finally, Figure 9 shows the histograms for Model $AB_1$. It is clear that at the final snapshot, both the ILR and the vILR are less populated than in Model C. The figure also suggests that the number of orbits that crossed the vILR and the mean $z_{max}$ around the vILR are smaller than in the control Model C, in agreement with the weakness of the BP/X bulge in this simulation.

In summary, these figures confirm the qualitative inferences drawn from the quasiaxisymmetric approximation in Figure 7, particularly the early development of the ILR and the vILR in Model $BF_0$, and that the effect of the ILR and vILR sweeping outward in the disk is to capture a significantly greater fraction of orbits into these resonances, strengthening the bar and causing greater levitation of orbits to high $|z|$ where they support the BP/X bulge.

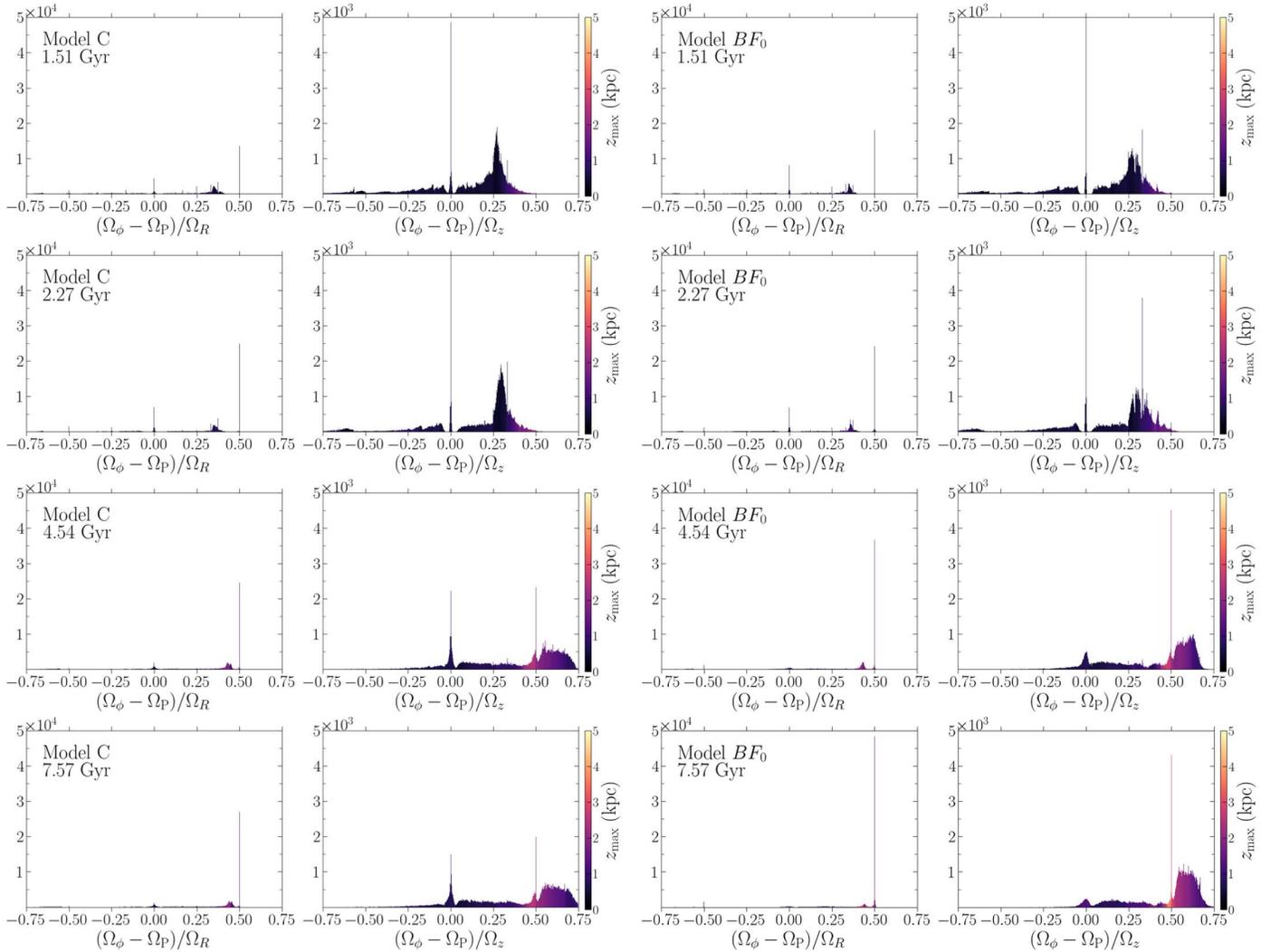

**Figure 8.** Histograms of frequency ratios for the same particles at different times (rows). The two left columns refer to Model *C*, and the two right columns refer to model $BF_0$. The corotation ($(\Omega_\phi - \Omega_P) = 0$), the ILR ($(\Omega_\phi - \Omega_P)/\Omega_R = 0.5$), and the vILR ($(\Omega_\phi - \Omega_P)/\Omega_z = 0.5$) are strongly populated.

To further understand why an early-growing SMBH (e.g., in Model $BF_0$) can strengthen both the bar and the BP/X bulge, while the SMBH grown after bar buckling weakens a steady-state bar, and the specific resonant orbit families that support the BP/X shape, we now examine frequency maps in Cartesian coordinates.

### 5.4. Frequency Analysis in Cartesian Coordinates and the Behavior of Resonant Bar Orbits

Valluri et al. (2016) showed that in frequency maps in Cartesian coordinates, for orbits integrated in the rotating frame of the bar, bar-supporting orbits are largely clustered in a cloud with $0.45 < \Omega_x/\Omega_z < 0.75$ and $0.65 < \Omega_y/\Omega_z < 0.95$, approximately the same region occupied by box orbits in stationary (nonrotating) triaxial potentials. These orbits (in the frame rotating with the bar) primarily originate from bifurcations of the linear long-axis orbit, which is the parent orbit of the "box" orbit family in stationary triaxial potentials and are referred to as the x1-orbit family in bars (Valluri et al. 2016). The BP/X-shaped bulge is associated with several families of resonant orbits, although it is not populated exclusively by resonant orbits (Abbott et al. 2017).

The introduction of a central point mass (representing an SMBH) in a (static) triaxial potential causes an increase in the number of resonances populated by box-like orbits (Valluri & Merritt 1998). As the mass of the SMBH increases, some of the resonances on the frequency map become thicker (due to increased resonant capture) while others are broken up or "fractured" due to the increased overlap of resonances. It has been well known since the early work of Chirikov (1979) that increasing the strength of a perturbation in a potential increases the number of resonances and that resonant overlap is an important factor driving the increase in the fraction of chaotic orbits. These chaotic orbits undergo mixing that drives the potential to a new dynamical equilibrium with fewer chaotic orbits (Merritt & Valluri 1998).

In Figure 10 we show frequency maps for Models *C*, $AB_1$, and $BF_0$ at $t = 4.54$ Gyr and $t = 7.57$ Gyr constructed from the integration of the same sets of $10^5$ randomly selected star particles considered in Figure 8. The points on the frequency map are colored by the logarithm (base 10) of the pericenter radius of each orbit (over the integration time of 100 orbital periods).

The top row of Figure 10 shows that the number and strength of most resonances in the control Model *C* do not change significantly between $t = 4.54$ Gyr and $t = 7.57$ Gyr (during which time the bar strength remains almost constant). In Model $AB_1$ (middle row) the frequency map at $t = 4.54$ Gyr (only 0.75 Gyr after the SMBH was introduced) already shows some differences from Model $C$, with several new weak resonance lines and increased clustering/scattering around the intersections of resonances. In addition, some strong resonances (e.g., $(1, -2, 1)$ and $(3, -5, 2)$) appear "broken up" or "fractured" at the places where they intersect other resonances. By $t = 7.57$ Gyr it is clear from the colors of the points that all of the orbits in Model $AB_1$ have significantly higher pericenter radii than they do in Model $C$ or even in the previous snapshot of Model $AB_1$ at $t = 4.54$ Gyr. Furthermore, as can be seen in Figure 12 at $t = 7.57$ Gyr, certain resonances, e.g., $(1, -3, 2)$, $(1, -2, 1)$, and $(2, 0, -1)$ in Model $AB_1$ are depopulated relative to their appearance at $t = 4.54$ Gyr.

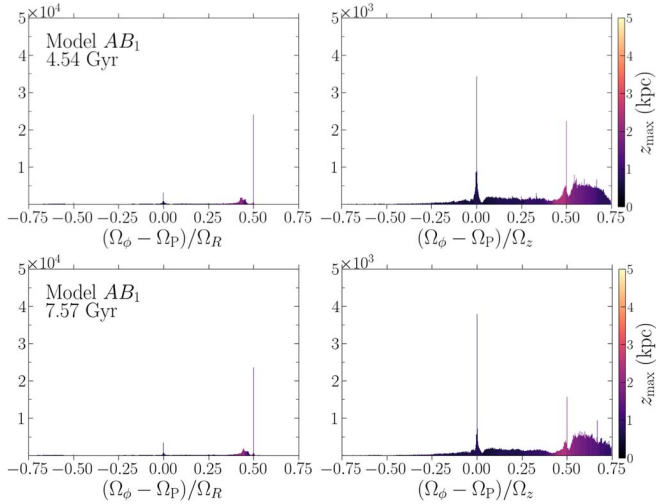

**Figure 9.** Similar to Figure 8, but for the Model $AB_1$ where the SMBH is introduced late, after bar buckling. The ILR and vILR are less populated than in Model $C$ in the final snapshot.

The bottom row shows frequency maps for Model $BF_0$ (SMBH introduced at $t = 0$). Recall that in this model the bar formed and grew around the fully grown preexisting SMBH. The frequency maps show that some resonances, e.g., $(1, -3, -2)$, $(2, 0, -1)$ are more strongly populated than in Model $C$ and some additional resonances (e.g., $(3, 0, -5)$, $(0, -1, 1)$) have appeared. However, some strong resonances in the top two rows (e.g., $(1, -2, 1)$ and $(3, -5, 2)$) are depopulated.

The overall structure of the frequency map in Model $BF_0$ is also quite different from that in the top two rows, reflecting the different orbit populations in this model. Since the bar forms and grows around a fully grown SMBH, the growing bar captures stars that then populate orbit families that are not destabilized by the presence of the central SMBH. The frequency map for Model $BF_0$ at $t = 7.57$ suggests that the bar-supporting resonant orbit families are populated in a different manner than in Model $C$. Additionally in Model $BF_0$ at $t = 7.57$ Gyr the diagonal resonance line ($(1, -1, 0)$; mostly occupied by disk stars) is much thinner than in the other panels—further evidence that the bar, having captured a significantly larger fraction of disk orbits, is much stronger in this model than in any other.

Although we do not show it, the Cartesian frequency map for Model $BB_1$ at $t = 7.57$ Gyr appears intermediate between the maps for Models $C$ and $BF_0$. Although the bar has already begun to form prior to SMBH introduction, its continued growth allows the stars to populate stable orbits.

Figure 11 shows histograms of the spherical pericenter radius $\log(r_{\rm per})$ for the $10^5$ orbits shown in the frequency maps in Figure 10. At $t = 4.54$ (left), Models $C$ and $AB_1$ differ very little in the distribution of $\log(r_{\rm per})$, and all three models show a distinct peak of orbits at very small radii $\log(r_{\rm per}/{\rm kpc}) < -1$. Note that this range of radii is comparable to the nominal $r_{\rm BH} \sim 0.1$–0.175 kpc of the SMBH in Model $AB_1$ and Model $BF_0$. It is important to point out that Model $BF_0$ (with the SMBH grown at $t = 0$) has a slightly higher fraction of orbits at the smallest $r_{\rm per}$ than even Model $C$ (at $t = 4.54$), which is evidence that a growing bar has absolutely no difficulty in finding ways to populate stable orbit families that pass quite close to the SMBH. Furthermore, the peak at the smallest $r_{\rm per}$ lies well inside the SMBH sphere of influence in Model $BF_0$ of 0.175 kpc.

By $t = 7.57$ Gyr, the peak in $r_{\rm per}/{\rm kpc} < -0.75$ has decreased slightly, consistent with the expectation that these orbits were scattered by the SMBH to large radii. Despite the fact that Model $BF_0$ has an SMBH of exactly the same mass as $AB_1$, the peak at the smallest radii is more highly populated at $t = 7.57$ Gyr, and has significantly more stars in this innermost region than Model $C$ (which has no SMBH).

The picture that emerges from the preceding analysis is that in Model $AB_1$, the SMBH that was introduced after the bar formed and reached equilibrium, scatters orbits with $\log(r_{\rm per}/{\rm kpc}) < -0.75$ to larger pericenter radii. Presumably since many of these were bar-supporting, their scattering weakened the bar and, because the bar potential is no longer growing at this point, stars cannot be captured onto alternative stable bar-supporting orbits. However, when the bar forms and/or grows around the SMBH (e.g., in Model $BF_0$), scattered orbits as well as newly captured orbits succeed in populating alternative centrophobic resonant orbit families that are bar-supporting.

The quantitative differences in the occupancy of resonant bar-supporting orbits in Models $C$, $AB_1$, and $BF_0$ can be seen in Figure 12, which shows the number of orbits with frequency ratios within $5 \times 10^{-3}$ of the major resonances that appeared in the frequency maps in Figure 10. Already at $t = 4.54$ Gyr, we see that the occupancy of some resonances (e.g., $(1, -3, 2)$, $(2, 0, -1)$) is significantly greater in Model $BF_0$ compared to Models $C$, $AB_1$. Other resonances such as $(1, -2, 1)$, $(3, -5, 2)$, and $(3, 0, -2)$ are depopulated. At $t = 4.54$ Gyr, the differences between Model $C$ and $AB_1$ are fairly small, but we begin to see changes in the population of resonant orbits. By the end of the simulation (right panel), we see that in Model $AB_1$ the number of orbits associated with several resonances ($(1, -3, 2)$, $(1, -2, 1)$, $(3, 0, -5)$, and $(2, 0, -1)$) is reduced relative to the numbers in Model $C$ (although one resonance is more strongly populated $(3, -5, 2)$, which interestingly is the resonance least populated in Model $BF_0$). Figure 11 indicates that introduction of the SMBH in Model $AB_1$ weakens the bar, primarily by decreasing the population of many bar-supporting resonances with the smallest pericenter radius. Because the bar is no longer growing after the introduction of the SMBH in Model $AB_1$ (and the ILR and vILR do not move in radius), there is no active mechanism by which the bar can adapt by capturing stars onto orbits that are stable in the presence of the SMBH. In contrast we saw in Figure 3 that if the bar amplitude continues to grow after buckling, then introducing the SMBH later may have little or no effect on the final bar strength.

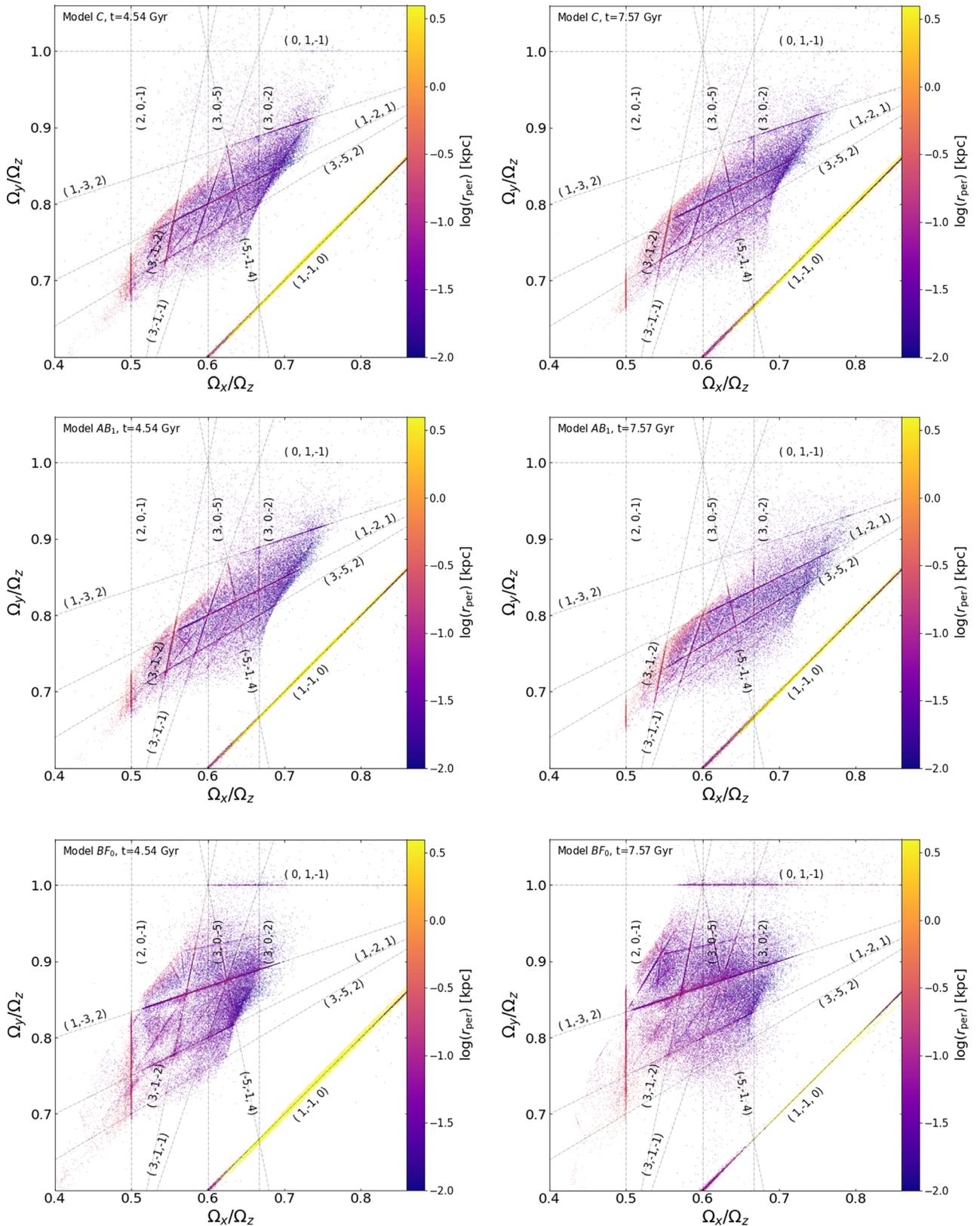

**Figure 10.** Frequency maps in Cartesian coordinates for the same set of $10^5$ orbits at $t = 4.54$ Gyr and $t = 7.57$ Gyr for Model $C$ (top row), Model $AB_1$ (middle row), and Model $BF_0$ (bottom row). Orbits in each panel are colored by the logarithm of their spherical pericenter radius $r_{\rm per}$ in that snapshot. Several resonances are marked in each panel as $(l, n, m)$ where the integers are coefficients of resonant conditions $l\Omega_x + m\Omega_y + n\Omega_z = 0$. The orbits clustered around the diagonal resonance line ((1, −1, 0)) at the bottom-right corner of each panel are primarily associated with the disk rather than the bar (see Valluri et al. 2016, for exceptions).

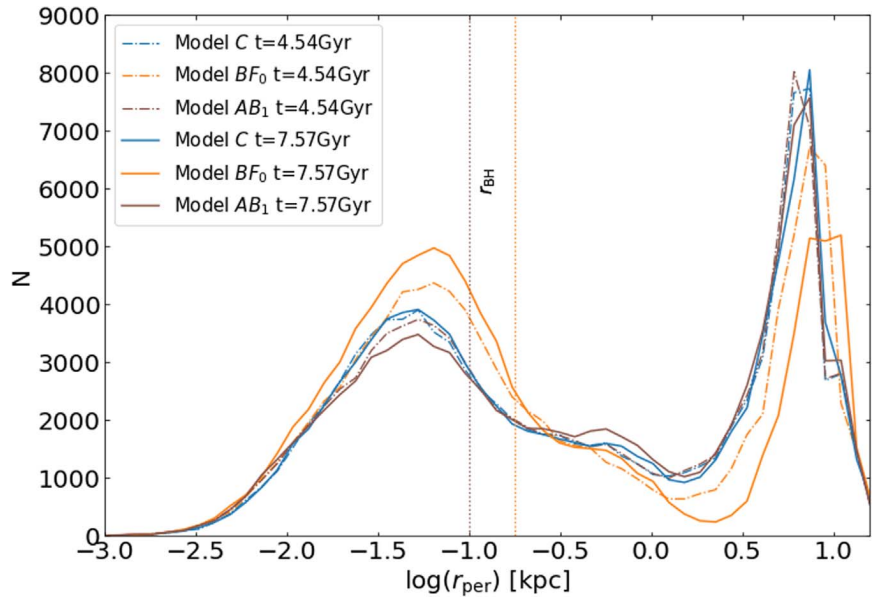

**Figure 11.** Histograms of the logarithm of the spherical pericenter radius in kiloparsecs for Models $C$, $BF_0$, and $AB_1$ at two different times (dashed–dotted lines are for $t = 4.54$ Gyr, solid lines are for $t = 7.57$ Gyr). At $t = 4.54$ Gyr (only 0.75 Gyr after the SMBH has been introduced), Model $AB_1$ differs only slightly from Model $C$. All three models have a significant peak at $\log(r_{\rm per}/{\rm kpc}) < -1$. By the end of the simulation, the peak in Model $AB_1$ at small $r_{\rm per}$ has decreased. In contrast, Model $C$ shows no significant change in the distribution, while in Model $BF_0$, the peak has significantly increased despite the presence of the massive black hole. The approximate location of the sphere of influence $r_{\rm BH}$ of the SMBH is shown in $BF_0$ and $AB_1$ by the vertical dotted lines, clearly indicating that orbits associated with the central peak in model $BF_0$ have $r_{\rm per}$ well within the SMBH's sphere of influence.

We note in passing that we also analyzed orbital frequency drift $\Delta_\Omega$ (Equation (14)) of all $10^5$ orbits in several models at $t = 7.57$ Gyr. Interestingly we find no difference in the distributions of this measure of chaoticity, and all models have a similar (small) fraction of chaotic orbits. This implies that although stars in Model $AB_1$ are scattered by the SMBH, by $t = 7.57$ Gyr they have successfully populated stable regular orbits (primarily at larger distances from the center).

## 6. Discussion

### 6.1. Are the Effects of an Early- versus Late-growing SMBH Observable?

The limited suite of simulations presented (as well as others we briefly discuss in Appendix A) show that if a bar reaches a steady state after buckling, then an SMBH that grows before buckling results in a greater $m = 2$ amplitude as well as a stronger BP/X shape (characterized by larger values of $R_{\rm pea}$ and $h_{\rm pea}$) than the control simulation without an SMBH or the simulation in which the SMBH grows after the bar reaches a steady state. However, we also saw from our scrambled control models that if the bar continues to grow after buckling, the late introduction of an SMBH may have little or no effect. We now ask if the early versus late growth of an SMBH can be inferred from observations of a galaxy at a single time, for instance from line-of-sight kinematics in external galaxies.

Sellwood & Gerhard (2020) argued that it is possible to distinguish between bars that buckle and cases where a BP/X bulge formed purely via resonant capture (no buckling) by observing differences in the $h_4$ component of the Gauss–Hermite expansion of the line-of-sight velocity distribution in the two cases. Restricting their measurement to observations of a binned vertical region $|z| < 200$ pc, with the disk viewed face-on, they find that as a result of buckling, $h_4$ becomes strongly negative as previously noted (Debattista et al. 2005). Sellwood & Gerhard (2020) found for "the bar that did not buckle," that $h_4$ remained positive or became more strongly positive. However, they found that the negative $h_4$ appeared to trend to 0 with time, and any systematic differences between cases became lost to noise by $|z| \gtrsim 500$ pc.

With the goal of searching for kinematic evidence for early-/late-growing SMBHs, we analyzed 2D kinematic maps (mimicking the Voronoi binned kinematics typically obtained for external galaxies) generated from our $N$-body snapshots both at early and late times in the evolution. Examples of kinematics maps are presented for four galaxies at the final snapshot ($t = 7.57$ Gyr) for both the face-on (Figure 13) and edge-on (Figure 14) orientations.

Unlike the case in Sellwood & Gerhard (2020), all of our models undergo buckling, although the magnitude of buckling varies and is dependent on when the SMBH is introduced. Therefore, we do not see a clear kinematic difference between the systems in which resonant trapping into the vILR played a stronger role (e.g., Models $BF_0$, $BB_1$), and models that buckled more strongly (Models $C$, $AB_1$). As previously noted (Debattista et al. 2005; Sellwood & Gerhard 2020), the face-on kinematic maps show strongly negative $h_4$ values in the central region, changing to positive $h_4$ values at the ends of the bar. In Figure 13 (face-on) we see that since the bar in Model $BF_0$ is much longer than in Model $C$, the distance from the center at which $h_4$ becomes positive is at or near the edge of the nominal field of view in our kinematic maps, but the behavior is otherwise similar. In Figure 14 (disk edge-on, bar side-on), we see in Model $C$ a positive $h_4$ in a central ring with two positive $h_4$ extensions in the disk plane. In Models $BF_0$ and $BB_1$, the ring-like $h_4$ region is broken up and now shows a quadrupolar structure. Although we do not show Models $BF_0$ and $BB_1$ at earlier times in their evolution when the BP/X shape was weaker, both of these models showed the same type of complete central ring-like structure with positive $h_4$, which appeared shortly after buckling (as in Model $C$), but changed to the quadrupolar structure as the BP/X shape grew stronger. We therefore believe that these differences only reflect the fact that the models with the quadrupolar $h_4$ have significantly stronger BP/X bulges than the models with ring-like $h_4$ structures. We also see that the $h_3$ bimodality along the bar is stronger in Model $C$ than in the models with early-growing SMBHs; but that too appears to be a consequence of the differences in the strength of the BP/X shape. It is outside the scope of this paper to determine exactly which features of orbital kinematics give rise to the different structures in $h_3$, $h_4$, and if signatures of evolutionary history (early versus late SMBH growth) may be extracted from a more detailed kinematic analysis.

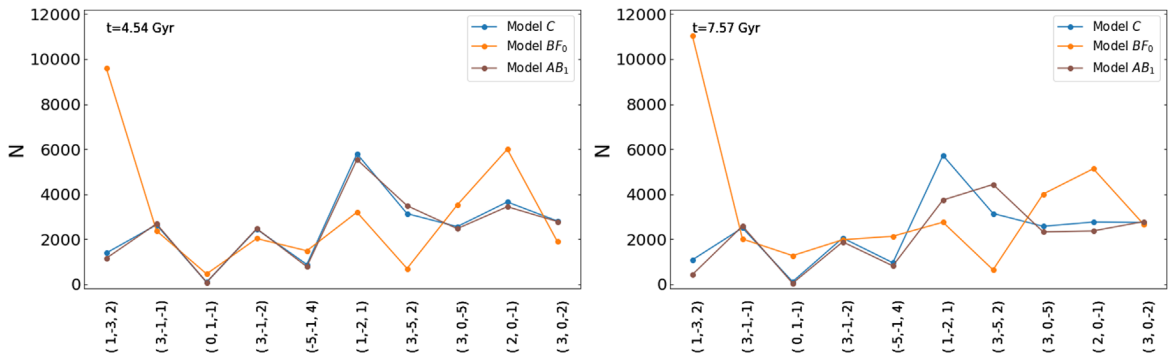

**Figure 12.** Number of orbits associated with important resonances seen in Figure 10. Orbits that lie with frequency ratios within $5 \times 10^{-3}$ of each of these resonance lines are shown at two snapshots as indicated.

### 6.2. Implications for the Offset of Bars from the $M_{BH}$–$\sigma$ Relation

As discussed in Section 1, it has been known for a long time that the tight $M_{\rm BH}$–$\sigma$ relation observed in elliptical galaxies and classical bulges does not appear to hold in the so-called pseudobulges (with low Sérsic index $n$ or exponential density profiles) or in BP/X-shaped bulges, both of which are thought to have formed by secular evolution. While the central velocity dispersion $\sigma$ in pseudobulges shows little correlation with $M_{\rm BH}$ (e.g., Kormendy et al. 2011), $\sigma$ may be systematically higher in barred spiral galaxies than in spirals with classical bulges (e.g., Graham 2008; Hartmann et al. 2014). The reason for the increase in scatter in pseudobulges and possible offset in barred spirals is still hotly debated.

The simulations presented in this paper have a fixed $M_{\rm BH}$ for all of the models, but the strength, and therefore mass, of the BP/X structure parameters $R_{\rm pea}$ and $h_{\rm pea}$ are correlated with $m=2$ bar amplitude. Also, as shown in Figure 6, both the radial and vertical central stellar velocity dispersions ($\sigma_R$, $\sigma_z$) are higher in models $BF_0$ and $BB_1$ than in Model $AB_1$, despite the latter having an SMBH of exactly the same mass. Consequently, our simulations have resulted in galaxies with the same $M_{\rm BH}$ but with BP/X bulges of different masses and different central velocity dispersions. We find that the average 3D velocity dispersion within $r = 1.5$ kpc in Model $AB_1$ at late times is 135 km s$^{-1}$ while in Model $BF_0$ it is 144 km s$^{-1}$, which is ∼7% larger for the same $M_{\rm BH}$. Furthermore, the average 3D velocity dispersion of Model $C$ at this time is 134 km s$^{-1}$, nearly identical to Model $AB_1$, despite not having an SMBH.

This supports previous arguments (e.g., Graham 2008; Brown et al. 2013; Hartmann et al. 2014) that the scatter in the $M_{\rm BH}$–$\sigma$ relation for barred galaxies is due to differences in bar strengths. Furthermore, quantities such as the half-light radius (within which $\sigma$ is generally computed) are harder to define when dealing with a BP/X-shaped structure rather than a classical bulge or elliptical galaxy since it is not well described by an ellipsoidal light distribution.

### 6.3. Improving upon N-body Results

The pure $N$-body simulations studied in this paper showed that the growth of an SMBH early in the life of a disk galaxy dramatically alters the formation and evolution of the bar. The SMBHs in these simulations were grown in an ad hoc manner (similar to previous $N$-body simulations of SMBH growth in bars), and therefore the simulations did not conserve mass, as one would expect in the case of a real galaxy. Future hydrodynamical simulations including star formation and more realistic prescriptions for SMBH accretion should be carried out to assess how general the results of our work are. In particular, since gas "cools" the disk by producing young stars on nearly circular orbits, disks with gas tend to have lower velocity dispersion. Therefore, simulations with gas and star formation may show less suppression of bar buckling. On the other hand, gas being collisional and dissipative is more likely to experience loss of angular momentum due to shocks in gas arising from their noncircular motions around the bar. Angular momentum transport by the bar could increase the growth rate of the SMBH and somewhat alter the evolution we have observed here. The most important point raised by our experiments is that, contrary to popular belief, SMBHs do not always weaken/destroy bars, and in fact bars have no difficulty forming around fully grown SMBHs or growing co-evally with SMBHs. In general we find that the presence of the SMBH in a growing bar, far from weakening the bar, may actually strengthen it by causing vertical and radial heating that strengthen both the ILR and vILR.

### 7. Summary and Conclusion

We have evolved $N$-body simulations of disk galaxies prone to bar formation, and have introduced and grown an SMBH at various times during the formation and evolution of the bar. In all cases, the SMBH's final mass and growth rate are held fixed, and only the time of introduction is varied. We obtain a number of results that lead to new insights into the formation of bars, factors that affect their buckling, and the formation and growth of BP/X-shaped bulges. We use a new framework for quantifying the strength of the BP/X-shaped bulge (from

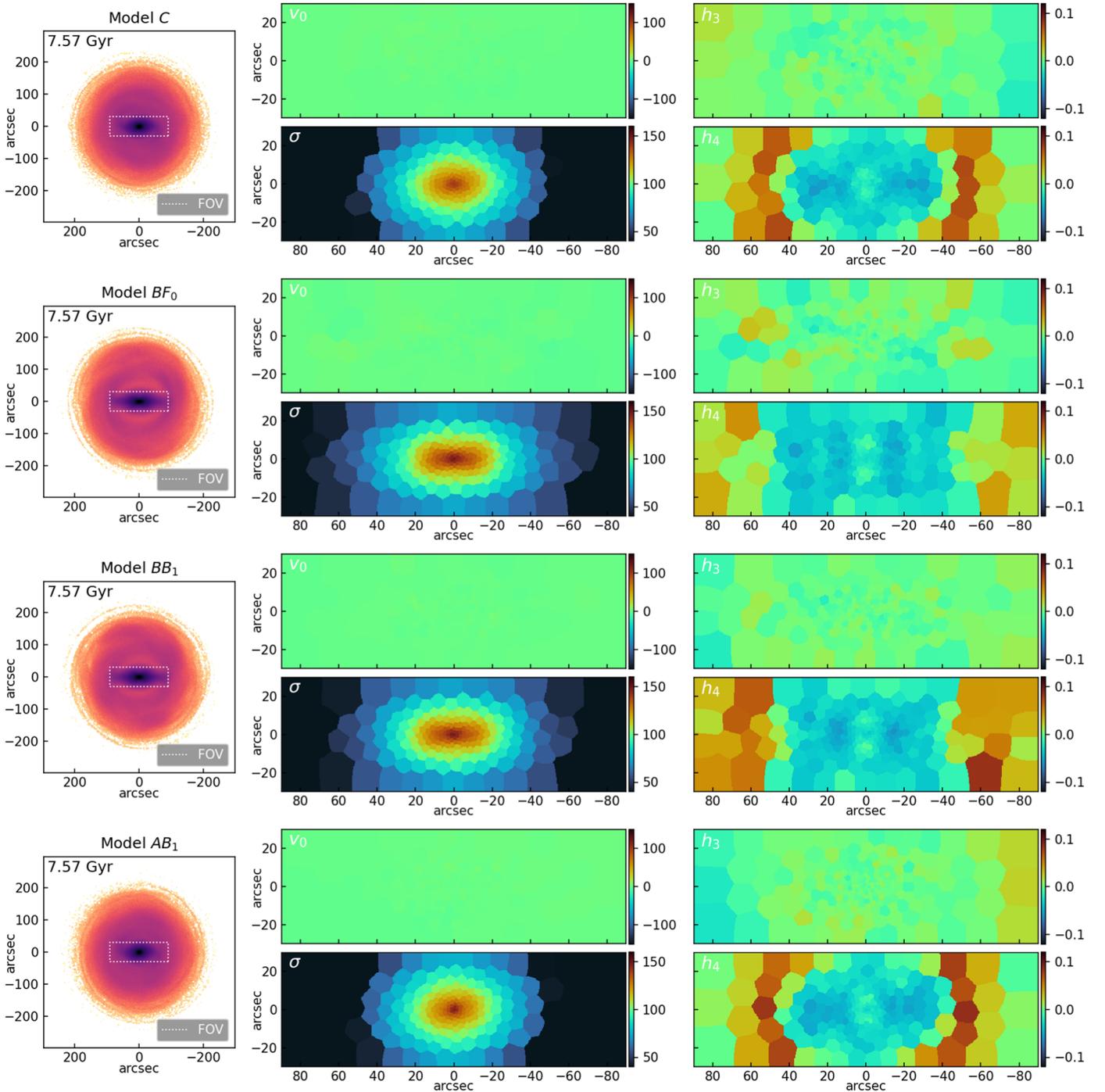

**Figure 13.** Mock IFU kinematic data generated from the final snapshot of Model $C$, $BF_0$, $BB_1$, and $AB_1$ (top to bottom), with the disk face-on and the bar aligned along the $x$-axis. From left to right: projected surface density and depiction of the kinematic field of view, line-of-sight velocity $v$, velocity dispersion $\sigma$, Gauss–Hermite coefficients $h_3$ and $h_4$ (as labeled).

Dattathri et al. 2023) to measure the strength/prominence of the BP/X-shaped bulge. We list our main findings below.

1. In all simulations in which an SMBH was introduced prior to bar formation or while the bar was still actively growing, bar amplitude is never decreased. In nearly every case, the bar amplitude relative to the control model without an SMBH is notably increased. The only cases in which the bar amplitude is decreased by the introduction of the SMBH is when the SMBH was introduced after the bar had reached a steady-state amplitude, consistent with previous findings (see Figure 1). In cases where the bar amplitude continues to increase after buckling, the late introduction of the SMBH has little or no effect (see Figure 3).

2. Two new parameters, $R_{\rm pea}$ and $h_{\rm pea}$, that characterize the strength of the BP/X shape, are found to be strongly correlated with the $m = 2$ bar amplitude (see Figures 4 and 5). If these correlations are found to hold for hydrodynamical simulations, it implies that the strength of a bar in an edge-on disk can be inferred from these observationally determinable quantities.

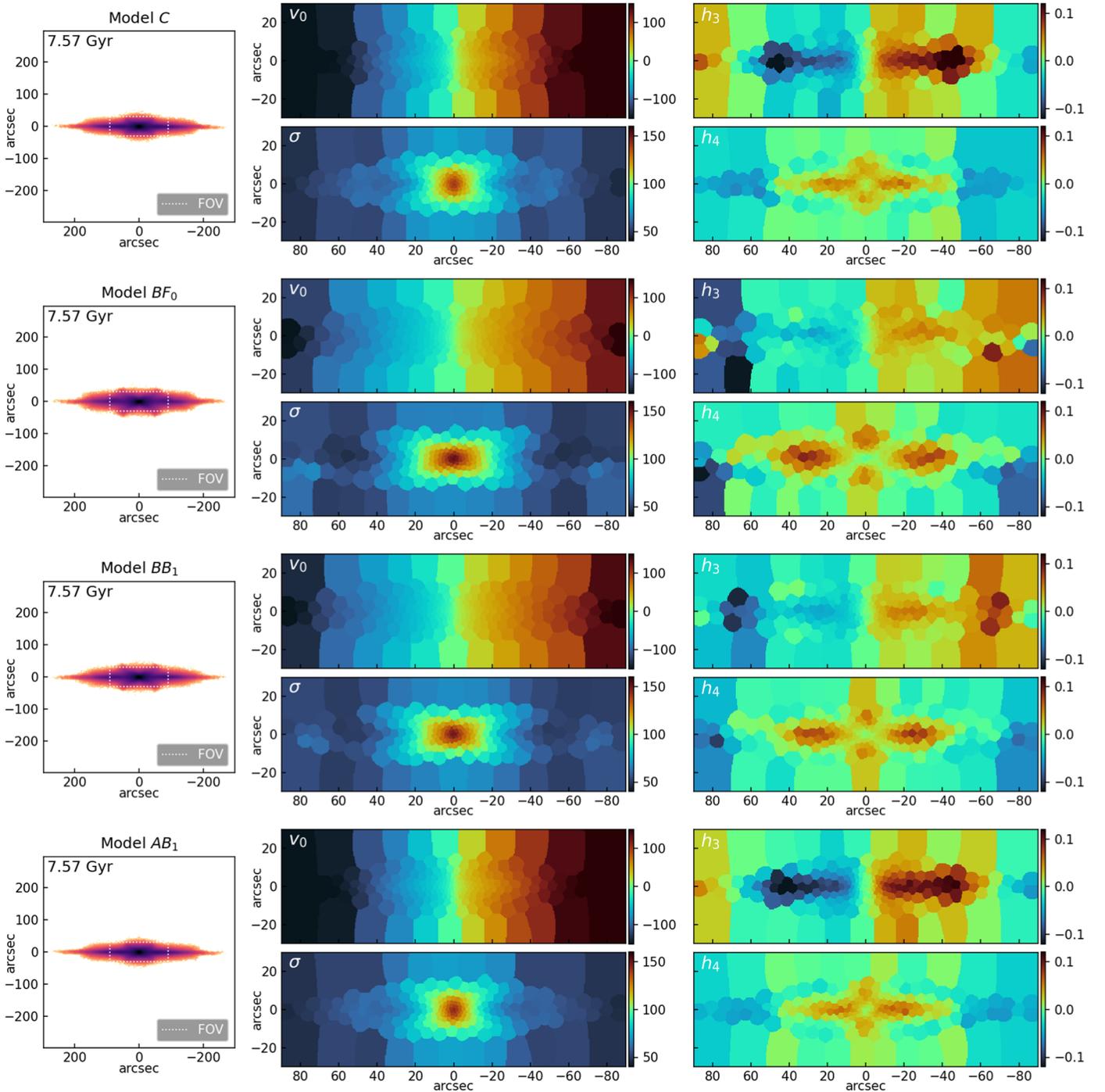

Figure 14. Same as Figure 13 with galaxies viewed edge-on and bar viewed side-on.

3. The introduction of an SMBH prior to buckling partially suppresses the buckling event either decreasing its magnitude, or delaying it (allowing the bar to grow stronger before eventually buckling), or both. This partial suppression of bar buckling is primarily a result of the increased vertical heating of the inner disk due to the presence of the SMBH. This decreases the velocity anisotropy (raises $\sigma_z/\sigma_R \gtrsim 0.6$), thereby making the bar less vulnerable to buckling (see Figures 6 and 15).

4. Vertical heating of the inner disk due to the SMBH causes a decrease in the vertical oscillation frequency $\nu$, and this results in the appearance of the vILR earlier than in the models without an SMBH. An SMBH introduced late (after the bar has buckled or reached steady state) causes little or no heating to an already vertically hot disk. We show that, in models with an SMBH, the ILR and vILR appear early (prior to bar buckling) and rapidly move outward in radius. The outward movement of these resonances (resonant sweeping) allows them to trap a larger fraction of orbits increasing the strength of the bar and the BP/X shape. Orbits that cross the vILR are raised to greater heights above the disk, contributing to a stronger BP/X shape. In models with a steady-state bar, the ILR and vILR move very little in radius after their

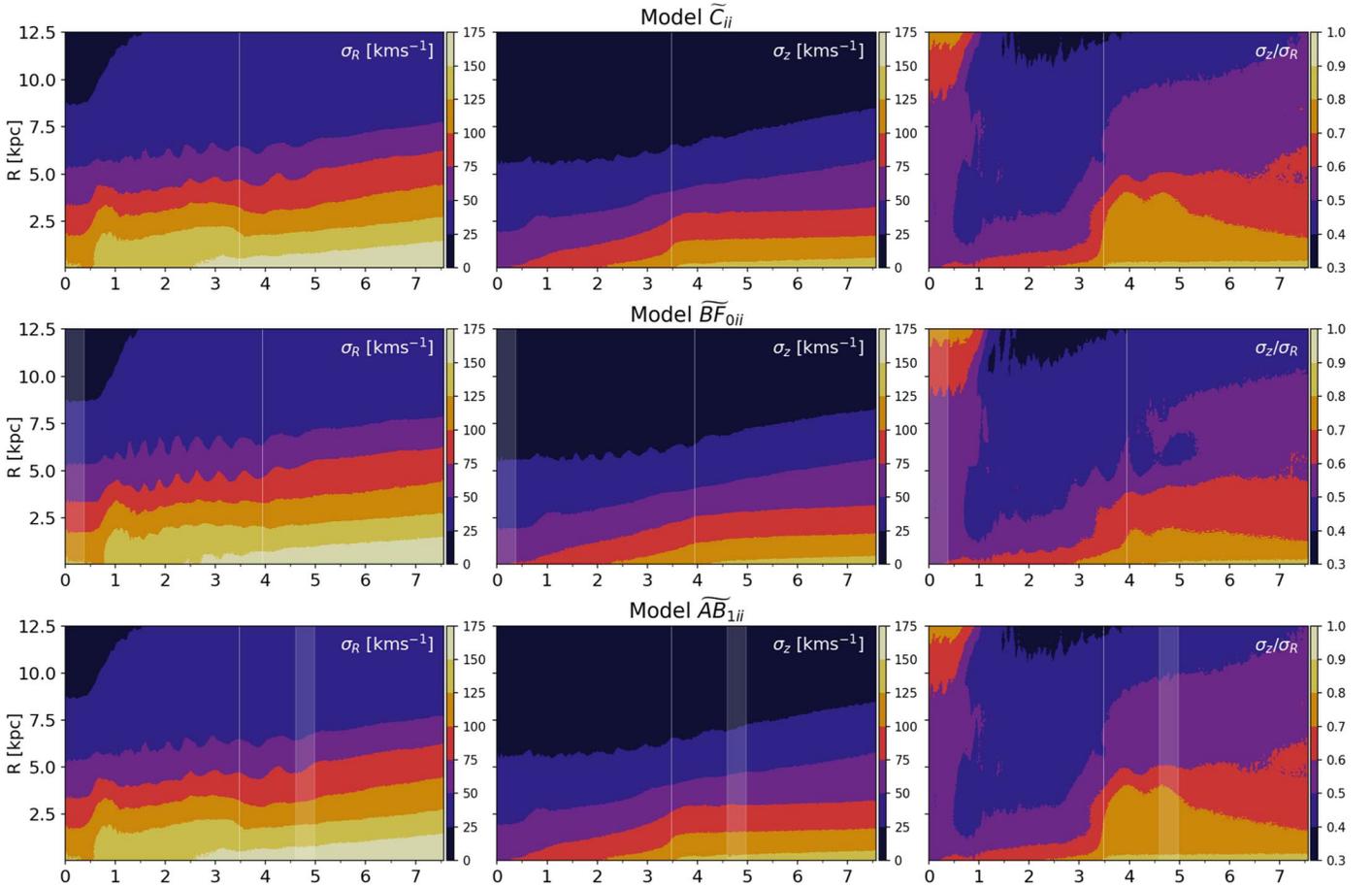

**Figure 15.** Similar to Figure 6 for control model $\widetilde{C}_{ii}$ and its $BF_0$ and $AB_1$ analogs.

   appearance, therefore limiting their ability to trap orbits
   and cause the bar to grow (see Figures 7, 8, and 9).
5. Using frequency analysis in Cartesian coordinates, we
   show that the specific bar-supporting resonant orbits in
   the bar depend on whether the bar formation and growth
   occurred in the presence or absence of an SMBH. We
   show that when an SMBH is grown after the bar has
   already formed and reached equilibrium, it scatters bar
   orbits at small pericenter radii resulting in a significantly
   weaker bar. However, when the bar forms around a
   preexisting SMBH or continues to grow after the SMBH
   is introduced, the specific resonant orbit families
   occupied and how heavily they are populated differ from
   the case where the bar reaches steady state before the
   SMBH is introduced. Many orbit families with small
   pericenter radii, which are stable in the presence of the
   SMBH, potentially exist, and are heavily populated (see
   Figures 10, 11, and 12). It is clear that the ability to
   populate orbit families that strengthen the BP/X shape
   can even be enhanced when an SMBH is present during
   bar growth. We argue that this is because vertical heating
   by the SMBH allows more orbits to be captured by
   the vILR.
6. Kinematic maps of our simulated galaxies viewed face-on
   and edge-on showing line-of-sight velocity ($v$), velocity
   dispersion ($\sigma$), and Gauss–Hermite coefficients ($h_3$, $h_4$)
   do not reveal any obvious "smoking gun" differences
   between models with early-growing, late-growing, or no

SMBH that can be attributed to the presence (or not) of an SMBH. Instead we believe that the kinematic differences we report are dominated by differing BP/X strength of our models. Further studies of models with matched BP/X and bar strength with early- and late-growing SMBHs are needed to determine definitively whether a signature of early growth is imprinted in the kinematics (see Figures 13 and 14.)

The rich variety of surprising results presented in this paper show that contrary to the currently prevailing view that SMBHs weaken bars, the preexisting or coeval growth of an SMBH with stellar bars may enhance both their strength and their BP/X structure, once again demonstrating that SMBHs influence galaxy evolution in profound and unexpected ways. Quasars and AGNs powered by SMBHs have been observed in disk galaxies at redshifts above $z = 2$ (e.g., Schawinski et al. 2011). On the other hand, disk galaxies form bars significantly later: the fraction of spiral galaxies containing bars is about 20% at $z \sim 0.85$ compared to about 65% in the local Universe (e.g., Sheth et al. 2008). Very recently, JWST CEERS NIRCam images have shown evidence for bars in disk galaxies at $z \sim 1$–2.3 (Guo et al. 2023). Although the number (six) of JWST imaged galaxies in which bars have been identified so far is too small to assess the global fractions of bars in disks at these redshifts, it appears that bar instabilities can occur early, and once formed, bars probably survive until the present time. This supports the view that bar-driven secular processes may have operated in disk galaxies over the past 8–10 Gyr.

Mass-matched samples of disks galaxies with/without X-ray luminous AGNs show that the bar fraction is comparable or marginally higher in disks with AGNs than in inactive disks (out to $z \simeq 0.85$; Cisternas et al. 2015). It is therefore reasonable to assume that a significant fraction of the SMBHs in barred galaxies began to form prior to bar formation and have continued to grow and strengthen stellar bars. The results we have presented here have potentially transformed our view of the coevolution of bars and central SMBHs.

### Acknowledgments

We thank Eric Bell for valuable discussions throughout the course of this project. We thank Jerry Sellwood for making the GALAXY N-body simulation code publicly available and for his assistance in the initial phase of this project. We also thank Eugene Vasiliev for making the AGAMA code (extensively used in this paper) publicly available, for his assistance in its usage, and for perennially enriching discussions. We thank Rishabh Ranjan for assistance in finding an error in a previous version of Figure 11. Finally, we thank the referee for constructive feedback that improved this paper. V.W., M.V., L.B.e.S., and S.D. gratefully acknowledge funding from the National Science Foundation (grants NSF-AST-1515001, NSF-AST-2009122) and the Space Telescope Science Institute (grant JWST-ERS-01364.002-A).

*Software:* GALAXY (Sellwood 2014), AGAMA (Vasiliev 2019), NAFF (Valluri & Merritt 1998), numpy (Harris et al. 2020), scipy (Virtanen et al. 2020).

### Data Availability

Snapshots from select models and times are available at doi:10.5281/zenodo.8230972.

### Appendix A
### Additional Models and Validation Tests

In addition to the models described in the main body of this paper, we ran several models both with the same set of initial conditions but with different particle softening lengths, integration time steps, simulation grid parameters (number of radial, azimulthal, and polar cells), the use or not of "guard radii," and fixed and freely moving SMBHs. We also ran models with additional sets of initial conditions. While there are numerous minor differences between the various models (such as how quickly the bar buckled, whether it continued to grow after the first buckling, and whether it buckled twice), in no case did we find that an early-growing SMBH weakened a bar. In contrast, in cases where the SMBH was grown after the bar reached a (quasi)equilibrium state, the SMBH weakened the bar consistent with previous works. Therefore, we can confidently state that the results presented in the main body of our work are not numerical artifacts of a specific setup or a result of stochastic behavior.

In Model $C$ the bar does not evolve after buckling, and therefore it is straightforward to compare the effects of the black hole grown at various times relative to the time of buckling. For some of the other initial conditions we considered (e.g., $C'$, $\widetilde{C}_i$, $\widetilde{C}_{ii}$, discussed in Section 3.3), the bar in the control model continued to grow after buckling, and consequently the effects of the early-growing SMBH were less dramatic (although they still increased the bar strength in all cases relative to the case with no SMBH).

In particular, $\widetilde{C}_{ii}$ has the strongest bar of all of our models, and all of the models with SMBH run with these initial conditions had nearly identical final bar strengths. Figure 15 shows the radial and vertical velocity dispersions $\sigma_R$, $\sigma_z$ and their ratio $\sigma_z/\sigma_R$ as a function of time and radius for this suite of models (similar to Figure 6). Although the initial radial and vertical velocity dispersions and density profile of this set of initial conditions is identical to Model $C$, the disk is more rapidly heated both radially and vertically. This causes both the ILR and the vILR to be strengthened, and the more rapid increase in $\sigma_z/\sigma_R$ stabilizes the disk so that it buckles only weakly (see Figure 2). As can be seen in Figure 15, the introduction of the SMBH either early or later does have a small effect but not enough to change the course of evolution.

Below we briefly describe our validation tests with other simulation parameters and results of tests with some of the other initial conditions. These additional initial conditions are also from Debattista et al. (2020).

*Runs with SMBH held fixed and guard radii.* Shen & Sellwood (2004) introduced guard-radii—a series of nested radial zones immediately surrounding the SMBH in which they progressively decreased, by a factor of 2 from the outermost guard-radius inward, the time step on which the position and velocity of a particle were updated (but the potential is only recalculated at the base time step). The guard radii are designed to improve the evolution of orbits in the vicinity of the SMBH where the forces change more rapidly. In our initial tests we also held the SMBH fixed at the center and used guard radii. The effects of the early-growing SMBH in these simulations were even stronger than in the models shown in the main body of this paper. We were concerned that holding the SMBH fixed in place (and turning off the $m=1$ sectoral harmonic as recommended in GALAXY documentation, thus disallowing forces that would translate the disk away from the origin) was artificially increasing the strength of the $m=2$ (bar) mode. We therefore removed the constraint of holding the SMBH fixed at the center.

Although the SMBH can be allowed to move, while using guard radii, the code centers the guard radii around the center of the grid and does not allow it to move with the SMBH. This results in unphysical effects, especially when the SMBH moves outside the region occupied by the guard radii. As noted in the main body of the paper, we do periodically relocate the origin of the grid to follow the location of highest particle density, and the guard radii are by definition centered about this origin; however, this does not guarantee that the SMBH lies within the innermost guard radius. Since it was not possible to modify GALAXY to enable the center of the guard-radii to move with the SMBH (and avoid unphysical numerical effects), we entirely eliminated the use of guard radii and instead reduced our base time step by a factor of 2 in all of our runs. While this is more computationally costly (since the smaller time steps are used everywhere), it achieves sufficient resolution given our SMBH softening length to enable the SMBH to move freely. Our main result—that the early-growing SMBH strengthens rather than weakens the bar—is unaffected by the presence or absence of guard radii, fixing the SMBH at the origin or allowing it to move, and enabling or disabling the $m=1$ mode (although the magnitude of the strengthening depends slightly on these details).

*Run with a fully grown SMBH at $t=0$.* In one model (not shown) using a feature of GALAXY, we turned off all harmonics other than the $m = 0$ mode, forcing the model to remain axisymmetric while we grew the SMBH to full mass over the same duration of 50 dynamical times. This resulted in a strictly axisymmetric disk with a potential that is self-consistent with the fully grown SMBH (of mass $0.0014 M_{disk}$). After the SMBH was fully grown and the disk had reached equilibrium, we turned all modes back on and evolved the model for a further 7.5 Gyr. This disk formed a strong bar. Since this model was evolved with the SMBH held fixed in place and used guard radii, we do not present the results here but note that similar results have been obtained by W. Dehnen et al. (2023, private communication) using NEMO with the Gyrfalcon integrator. We therefore are confident that bars are quite capable of forming from axisymmetric disks around fully grown preexisting black holes (with masses ∼0.2% disk mass), and the presence of a fully grown SMBH does not prevent the formation of a large-scale bar. The later behavior of this model in terms of strengthening of the bar/bulge and suppression of buckling is very similar to the comparable case in which the SMBH was grown beginning at time 0.

*Run 5000 series.* This set of initial conditions was referred to as Model 1 in Debattista et al. (2020). All of the simulations in the Run 5000 series experienced strong buckling on a shorter timescale than models presented in the main body and showed no continued bar growth after buckling both for cases with and without an early-growing SMBH. However, the relative differences in bar strength and the BP/X bulge strength are still present but are less strong at late times than in the models described in the main body of this paper, and are thus not shown in detail.

*Run 6000 series.* Models of run 6000 series, evolved from initial conditions of Model 3 in Debattista et al. (2020) are a case in which the bar very quickly undergoes a small buckling event, then subsequently experiences strong buckling on a longer timescale and continues to grow thereafter in cases both with and without an early-growing SMBH. Since the bar strength in these models continues to increase in all cases, the bars then buckle at least once more. The long timescale and continued growth makes it difficult to predict the final outcomes of the SMBH introduction to this system even with a simulation time of 15 Gyr. This model suggests that the SMBH suppressing/delaying/weakening of the first strong buckling event (in addition to the early weak event) might lead to a subsequent buckling event (which primarily occurs in the outer bar at large R) that is stronger in amplitude although still delayed, appearing to result in a weaker bar relative to the case with no SMBH. However, the bar in both cases still continues to grow for some time after this buckling event. Further testing in systems that undergo repeated buckling is required to explore the SMBH role in such systems and to answer questions such as: would a third strong buckling event again show the SMBH model to have weaker buckling and result in stronger bar strength (a leap-frog effect with regard to buckling)? and what are the effects of SMBH introduced between two strong buckling events? This experiment suggests that further investigation is required to fully understand the interaction between the SMBH and the bar, but shows that this interaction is much more nuanced than previous generations of experiments implied.

## Appendix B
## Grid Parameters

Listed below is a full account of input parameters used to specify model evolution in GALAXY version 15.4, which makes up a simulation startup file. Unless otherwise noted, all keywords and values correspond to built-in commands, which are explained in the public documentation available at http://www.physics.rutgers.edu/galaxy/. All dimensional values here are entered in simulation units: the gravitational constant $G = 1$, scale radius is set to unit length $L = 1$ (some inputs further scaled by grid units: $L \times 1$ scale), disk mass is set to unit mass $M = 1$, dynamical velocity defined from the previous quantities is also $v_{dyn} = 1$, thus so is a unit of dynamical time $T_{dyn} = 1$. Scaling relations follow from specifying any two of the above in physical units, usually $L$ and $M$.

```
run no 2099                   # Internal model number for
                                model BF_0
grid type HYB 2               # Select 2 grids of different
                                types
1st type p3d
2nd type s3d
#
ncom 2                        # Active mass components
#component 1, the disk
disk t                        # Specify is disk
type UNKN                     # UNKN used to specify exter-
                                nally created component
mass 1
scale 1 1
dftype none 1.5               # Unused DF and dummy para-
                                meter (ext comp)
taper f                       # No taper (ext comp)
#component 2, the halo
disk f
type UNKN
mass 1
scale 1
dftype none
# Instructions for p3d
grid size 172 256 405         # Radial division, azimuthal
                                division, z-division
z spacing 0.1                 # In grid units
HASH 8                        # Highest active sectoral
                                harmonic
softl 0.2083                  # In grid units (shared by s3d)
sect 1                        # Impose no symmetry con-
                                straints (shared by s3d)
skip                          # Skip no harmonics (shared
                                by s3d)
# Instructions for s3d
grid size 400 800             # Radial divisions, Radius of
                                outer boundary (grid units)
lmax 8                        # Highest active spherical
                                harmonic
# Instructions for setrun
time step 0.005               # Units of T_dyn
zones 5                       # Number of t step zones
 2 1                          # Advance motion every N steps
                                outside radius R
 4 3
 8 5
 16 8
lscale 10                     # Set scale grid unit scale factor
                                —10 grid units per L
offgrid t f f f               # Flags for off grid particles
```

| (Continued) | |
|---|---|
| uqmass t | # particle unique mass (req as disk and dm differ) |
| cntd 48 | # recenter grids every 48 time steps |
| supplement f | # No supplemental forces added |
| # Instructions perturber | # Section omitted for cases with no SMBH |
| perturber GNRC 0.0014 0.0139 0.02 50.0 | # * significantly edited, see below |
| position 0.0 0.0 0.0 | # initial position |
| velocity 0.0 0.0 0.0 | # initial velocity |
| # further component info | |
| icmp 1 | # comp no |
| npar 6000000 | # particle no |
| z0in 1 | # initial thickness—dummy input (ext comp) |
| pgrd 1 | # primary grid assignment |
| start f f f | # startup flags (all f, ext comp) |
| end data for component 1 | |
| # | |
| icmp 2 | |
| npar 4000000 | |
| pgrd 2 | |
| start f f f | |
| end data for component 2 | |
| # GALAXY analysis flags | # Analysis using built-in tools** |
| analysis 800 | # Perform analysis and output snapshot*** every 800 steps |
| save | # Following keywords specify analysis to be performed |
| intg | |
| danl | |
| end of results file instructions | |
| end | |
| # | |
| last step 200000 | |

1. Inputs specifying perturber treatment have been significantly edited from the original GALAXY version 15.4 to facilitate evolution of rigid perturber mass. The first three inputs following the keyword are default parameters that specify a generic type perturber, which is used for user-defined perturber potentials. Following this is the mass and softening length in units of $M$ and grid units, respectively. The following two are bespoke additions that specify the initial mass fraction (of final mass of the perturber) and the growth period in $T_{\rm dyn}$.
2. All analysis presented in this paper was performed using methods external to GALAXY.
3. In the default prescription, snapshots are saved every 25 analysis steps. However this implementation has been edited to output a snapshot on every analysis step. This simplifies the input parameter.

## ORCID iDs

Vance Wheeler  https://orcid.org/0000-0003-4679-4435
Monica Valluri  https://orcid.org/0000-0002-6257-2341
Leandro Beraldo e Silva  https://orcid.org/0000-0002-0740-1507
Shashank Dattathri  https://orcid.org/0000-0002-7941-1149
Victor P. Debattista  https://orcid.org/0000-0001-7902-0116